\documentclass[aps,prd,epsf,showpacs,amsmath,amssymb,graphics,onecolumn,preprint,12pt]{revtex4}
\usepackage{graphicx}
\usepackage{dcolumn}
\usepackage{bm}
\usepackage{amsmath}
\usepackage{mathtools}
\usepackage{physics}

\newcommand{\be}{\begin{equation}}
\newcommand{\ee}{\end{equation}}
\newcommand{\ba}{\begin{eqnarray}}
\newcommand{\ea}{\end{eqnarray}}
\newcommand{\ban}{\begin{eqnarray*}}
\newcommand{\ean}{\end{eqnarray*}}

\begin{document}
\title{Gravitational lensing signature of matter distribution around Schwarzschild black hole}
\author{Harsha Miriam Reji and 
Mandar Patil\footnote{Electronic address: mandar@iitdh.ac.in}}

\affiliation{Indian Institute of Technology Dharwad, Dharwad, Karnataka 580011, India.}

\begin{abstract}

In this work we focus on the situation where significant amount of matter could be located close to the event horizon of the 
central black hole and how it affects the gravitational lensing signal.  We consider a simple toy model where matter is 
concentrated in the rather small region between the inner photon sphere associated with the mass of central black hole and 
outer photon sphere associated with the total mass outside. If no photon sphere is present inside the matter distribution, then 
effective potential displays an interesting trend with maxima at inner and outer photon sphere, with peak at inner photon sphere 
higher than that at outer photon sphere. In such a case we get three distinct set of infinitely many relativistic images and Einstein 
rings that occur due to the light rays that approach the black hole from distant source and get reflected back just outside 
the outer photon sphere, due to light rays that enter the the outer photon sphere slightly above the outer peak and get reflected off
the potential barrier inside the matter distribution and due to the light rays that get reflected just 
outside the inner photon sphere. This kind of pattern of images is quite unprecedented. We show that since relativistiv images are highly demagnified, only three images are prominiently visible from the point of observations in the presence of matter as opposed to only one prominent image in case of single isolated black hole and also compute the time delay between them. This provides a smoking gun signature of presence of matter lump around black hole.    
We further argue that if the mass of the black hole inferred from the observation of size of its shadow is less than the 
mass inferred from the motion of objects around it, it signals the presence of matter in the vicinity of black hole.

\end{abstract}


\maketitle

\section{Introduction}

Einstein proposed General Theory of Relativity in 1915, a theory of gravity which superseded the Newtonian theory 
proposed almost four hundred years ago. General Relativity came up with many new predictions of the phenomenon which were absent in the Newtonian theory of gravity. We shall focus on two such predictions in this paper, namely 
the black hole and gravitational lensing. Black holes are the objects endowed with a one-way surface known as event horizon which absorbs
all incoming matter and radiation and does not let anything go out. The first exact solution was proposed way back in 1917 which 
represents a spherically symmetric black hole parametrized by its mass. It is referred to as Schwarzschild black hole. 
Gravitational lensing is a phenomenon where light from a source gets bent as it passes by a gravitating object and thus affecting 
the location of the source as inferred by the observer \cite{lr}. Bending of the light around sun was observed during solar eclipse in 1919
and was consistent with the theoretical prediction based on general relativity. In this paper we study the gravitational relativistic gravitational lensing by a Schwarschild black hole surrounded by matter in its vicinity and contrast it with gravitational lensing by an isolated black hole. The lensing said to be relativistic if the deflection angle is more than $\frac{3}{2}\pi$ \cite{vir},\cite{vir2}.
We show that in the presence of matter the pattern of images is very peculiar and quite unprecedented.

As mentioned earlier first black hole solution was proposed way back in 1917 almost immediately after Einstein proposed General relativity \cite{Sch}.
But because of the esoteric nature of the object is represents, it was considered to be a mathematical artifact and it was believed 
that it does not represent a realistic object that can exist in the universe. The dynamical solution that represents gravitational collapse
of homogeneous dust, a pressure-less matter, that produced Schwarzschild black hole as an end-state was found in 1939 \cite{OS}. The black holes 
were taken seriously in 1960`s when extremely energetic phenomenon such as Quasars and Active Galactic Nuclei were discovered
and there was no other alternative explaination for production of large amount of energy on such a small timescales \cite{agn}. But now 
black holes lie at the heart of both observational astronomy as well as theoretical physics. We believe that there is a supermassive black hole 
at the center of almost every galaxy with the masses that range from million to billion times the mass of of sun \cite{smbh}. It is estimated that 
there are several million to billion black holes which have mass comparable to the mass of sun lurking around in our galaxy.
Yet another prediction of general relativity is gravitational waves which are tiny fluctuations of geometry of spacetime
that travel at speed of light. We have detected gravitational waves from the merger to two black holes, an astrophysical event where 
two black holes spiral around one another and eventually merge together to form a single black hole \cite{gw1}, \cite{gw2}. It is believed that the black hole forms
when the stars with large enough mass at the end of their life-cycle when they run out of fuel implode due to self-gravity and
eventually turn into black black hole \cite{collapse}. Thus black holes are everywhere and inevitable. Since gravity in their vicinity of black holes
is strong it offers us an opportunity to understand the General Relativity and its alternatives in strong field regime.

Gravitational lensing around stars such as sun is small with tiny bending angle which results into tiny displacement of image of the source \cite{wein}.
Things change drastically in the vicinity of black holes where gravity is strong and light can suffer a significant bending, the phenomenon
which is referred to as relativistic gravitational lensing \cite{vir}. 
Light can go around the black hole in the circular orbit which is called photon sphere that is located at $r=3M$ while event 
horizon is located at $r=2M$. Here we work in geometrical units where $c=G=1$. The amount of deflection suffered by light depends
on the minimum distance of approach or the radius at which light ray approaching the black hole turns back and starts its outward
journey. As the turning point gets closer to the photon sphere deflection angle goes on increasing and shows divergence as it approaches
the photon sphere. So, depending on the location of turning point we can have light rays that go around black hole once, twice, thrice and so on all the way 
upto infinite turns, which leads to the formation of infinitely many images of the same source. If this investigation we assume that 
the source is located almost exactly behind the black hole slightly off. If it is in turn exactly behind the source then the image is 
ring which is referred to as Einstein ring. We get infinitely many concentric rings. If the light ray were to turn back it must turn back
outside the photon sphere. If light ray enters  the photon sphere it will inevitably be engulfed by the black hole event horizon, essentially
because of which we get a dark patch in the middle where no images or rings can lie. It is referred to as the shadow of a black hole.
There are many investigations in the literature which deal with the gravitational lensing by different kind of black holes \cite{vir},\cite{bozza},\cite{rn},\cite{lqg}.
People also study gravitational lensing around more exotic objects such as naked singularities, wormholes etc \cite{ns1},\cite{ns2},\cite{ns3},\cite{ns4},\cite{ns5},\cite{wh1},\cite{wh2},\cite{wh3},\cite{wh4} where photon sphere may or may not be present. In this 
paper we focus our attention on black holes. Photon sphere which is the circular photon can be easily obtained by locating maximum of effective potential in Schwarschild spacetime. The concept of photon sphere in Schwarschild spacetime has been generelized to arbitary spacetimes \cite{Clau}. Since we consider essentially the Schwraschild photon spheres, the general definition should quite naturally reduce to the notion of photon spheres used in our paper. 

Black holes are compact objects. Event horizon of one solar mass black hole is of the size 3 km, an opposed to the size of the sun
which is around 0.7 million km. The radius of the event horizon of the black hole is proportional to its mass. Approximately the 
angle subtended by the black hole at earth is the size of the black hole over distance to the black hole which is
ridiculously small. Thus we need telescope with extremely large resolution in order to observe the black hole and the phenomenon
occurring in its vicinity. This is an impossible task with a single telescope. Thus we invoke the technique of interferometry where we 
combine the data from different telescopes widely separated so as to form a virtual telescope with large size, significantly
improving the sensitivity and resolution. The GRAVITY experiment combines four telescopes to form an optical telescope with effective 
diameter of around 100 m \cite{grv}. Whereas Event Horizon Telescope combines various radio telescopes across continents generating a telescope
with effective diameter comparable to the size of the earth. Event horizon Telescope has observed a black hole at the center of our
neighboring Andromeda galaxy \cite{eh}. So far the observation is consistent both General Relativity and black hole nature of central object.
But observations in future with better sensitivities of instrument involved, not only with light but also with gravitational waves,
neutrinos etc might throw surprises. Many investigations were done exploring possibility beyond simple Kerr-black hole nature of compact object at the center of andromeda galaxy \cite{sha1},\cite{sha2},\cite{sha3},\cite{sha4}.

In this investigation we study the gravitational lensing by a Schwarzschild black hole surrounded by spherically symmetric 
matter distribution. The black hole at the center is assumed to have mass $M_1$ and its event horizon is located at $r=2M_1$.
The photon sphere is located at $r_{ph1}=3M_1$. The mass outside is $M_2$ with $M_2>M_1$ and thus photon sphere associated with it is
located at $r_{ph2}=3M_2$. We assume that the mass distribution is located between two photon spheres. If mass lies between $r=r_1$ to $r=r_2$
with $r_2>r_1$, then we have $r_{ph1}<r_1<r_2<r_{ph2}$. With further assumptions on the mass distribution we get an 
effective potential for radial motion of light rays that displays an interesting trend. It admits two maxima, one at the inner
photon sphere $r=r_{ph1}$ and one at outer photon sphere $r=r_{ph2}$. The value of the potential at maximum is larger at inner photon
sphere. We get three set of infinitely many images and Einstein rings. One set is due to the light rays that get reflected back at the location outside the outer photon sphere. Second set of images is due to the light rays that enter the outer photon sphere just above 
the maximum and get reflected at in the region where mass is distributed. And third set of images is due to the light rays that 
get reflected back close to the inner photon sphere. This pattern of images is quite unprecedented and differs significantly from that 
of an isolated single Schwarzschild black hole. Relativistic images are highly demagnified as the number of turns around black hole increases. We show that in the presence of matter around black hole 
three images are quite prominently visible as compared just one prominent image in case of single isolated black hole, 
which would provide a smoking gun signature of distribution of matter close to the black hole. We also compute time delays between three image which would be relevant if the source is variable. Further the size of the shadow is dictated by the inner photon sphere, since peak in the effective potential is higher and thus inferred mass from the shadow will be $M_1$. Whereas the inferred mass from the motion of the objects in the vicinity will be outer mass $M_2$. Thus we can infer the presence of the matter close to the black hole from the mismatch of the inference from two different observations. The effect of matter distribution surrounding black hole on its shadow was investigated in \cite{Cunha},\cite{Hou},\cite{Haroon},\cite{Hou2},\cite{Konoplya} where different kinds of matter fields such as scalar field, perfect fluid dark matter, dark energy etc. were considered. Here the matter affects the spacetime geometry and thereby affecting the propogation of light. If the plasma is present around black hole it can also affect the propagation of light due to the electromagnetic interaction \cite{Abdu},\cite{Abdu2},\cite{Abdu3}. But rigorous analysis exploring effect of pattern of relativistic images and whether or not it displays interesting trend
is not explored so far. Here we consider a very simplistic toy model with minimal assumptions about the nature of matter and show that 
under the circumstances considered here we get very interesting and novel pattern of images due to the presence of multiple photon spheres. 

An investigation somewhat similar to this was carried out by one of us (MP) where we studied gravitational lensing by binary black holes.
We consider Majumdar-Papapetrou spacetime which represent two identical mass black holes in equilibrium \cite{bbh}. We focus our attention on 
the light rays that move on the plane midway between the two black holes. If the two black holes are close enough then the effective 
potential looks similar to the one in this paper except for the fact that inner photon sphere in absent and potential diverges as we 
go in the inward direction. Thus we get two set of relativistic images. One set is due to the light rays that get reflected back
just outside photon sphere and other set is due to the light rays that enter photon sphere just above the maximum and then get reflected 
black off the potential barrier. We had developed a formalism to calculate location of images in terms of effective potential \cite{bbh}.
We import and use those techniques here in this paper.

The organization of paper is as follows. In section II we discuss null geodesics in spherically symmetric spacetime and around Schwrzschild black hole. In section III we present the gravitational lensing formalism. In section IV we describe the system under consideration i.e. black hole with matter surrounding it. In section V we describe image formation due to the light rays that turn back outside the outer photon sphere. In section VI we discuss images due to the light rays that enter outer photon sphere just above the peak and get reflected inside matter. In section VII we discuss formation of images due to the light rays that get reflected outside inner photon sphere. In section VIII we discuss the shadow cast by the black hole with matter in the surrounding. In section IX we discuss the properties of most prominent images that would be relevant for observations. In section X we provide summary and concluding remarks. 

\section{Null geodesics in spherically symmetric spacetime and around Schwarzschild black hole}

In this section we review the null geodesics in static spherically symmetric spacetimes, in particular focusing on Schwarzscild metric which is the case of interest to us. We consider general static spherically symmetric spacetime. Its line element is given by 
\begin{equation}
 ds^2=-f(r)dt^2+\frac{dr^2}{f(r)}+r^2\left(d\theta^2+\sin^2{\theta}~ d\phi^2 \right)
 \label{gsph}
\end{equation}
In general the inverse of a coefficient of $dr^2$ could be different from coefficient of $dt^2$, but we focus on the case
where they are identical. We consider a light ray that follows geodesic motion. We can get equations describing geodesic by varying the Lagrangian $L$ given below 
\begin{equation}
 L=\frac{1}{2}g_{\mu\nu}\dot{x}^{\mu}\dot{x}^{\nu}=\frac{1}{2}\left( - f(r)\dot{t}^2+\frac{\dot{r}^2}{f(r)}+r^2\left(\dot{\theta}^2+\sin^2{\theta}~ \dot{\phi}^2 \right)\right) 
\label{lag}
 \end{equation}
where derivative is with respect to affine parameter $\lambda$ and $\dot{x}^{\mu}$ is the four-velocity of the light ray. Euler-Lagrange equation of motion are given by
\begin{equation}
\frac{d}{d\lambda}\left(\frac{\partial L}{\partial \dot{x}^{\alpha}}\right)=\frac{\partial L}{\partial x^{\alpha}}.
\label{el}
\end{equation}
They are supplemented with the normalization condition $L=0$ for null geodesic.
The $\theta-$component of Euler-Lagrange equation of motion turns out to be
\begin{equation}
\ddot{\theta}=\dot{\phi}^2 \sin{\theta}\cos{\theta}-2\frac{\dot{r}\dot{\theta}}{r}.
\end{equation}
So if at the initial moment $t=t_i$ particle is moving on the equatorial plane, i.e. $\theta(t_i)=\frac{\pi}{2}$ and 
$\dot{\theta}(t_i)=0$, then $\ddot{\theta}(t_i)=0$
and it implies that particle would continue to move on the equatorial plane.
We focus our attention on the light rays that move on the equatorial plane
and thus now onward $\theta$ will take a value $\theta=\frac{\pi}{2}$. This is not a loss of generality, since in spherically symmetric spacetime light will move on a plane passing through the center, which by relabeling the coordinates can be turned into the equatorial plane. Since none of the metric coefficients depend explicitly on time $t$ and azimuthal angle $\phi$, right hand side of $t-$component and $\phi-$component 
of Eq.\ref{el} will be zero and hence the quantities in the paranthesis on left hand side would be constants of motion, which 
are referred to as conserved energy $E$ and conserved angular momentum $L$.
From this consideration we get equations for the time and azimuthal components of four-velocity. Combining these equations
with Eq.\ref{lag} and the normalization condition $L=0$, we also get the equation for the radial component of velocity. We have 
\begin{eqnarray}
 &&\dot{t}=\frac{E}{f(r)} \nonumber \\
 &&\dot{\phi}=\frac{L}{r^2} \nonumber \\
 &&\dot{r}^2  +\frac{L^2}{r^2}f(r)= E^2. 
 \label{eq1}
\end{eqnarray}
We can write down the equations above with the reparametrization $\lambda \rightarrow \lambda /|L|$. We get
\begin{eqnarray}
  &&\dot{t}=\frac{1}{b f(r)} \nonumber \\
  &&\dot{\phi}=\pm \frac{1}{r^2} \nonumber \\
  &&\dot{r}^2+\frac{1}{r^2}f(r)= \frac{1}{b^2}. 
 \label{eq2}
\end{eqnarray}
The parameter $b$ which appears in the equations above is $b=\frac{|L|}{E}$ and called the impact parameter. Plus and minus
sign corresponds to the light rays that travel in counterclockwise and clockwise respectively along azimuthal direction.
We can rewrite the last equation as 
\begin{equation}
 \dot{r}^2+V(r)=\frac{1}{b^2}~~~~~\text{with} ~~~~~ V(r)=\frac{1}{r^2}f(r),
 \label{ve}
\end{equation}
where $V$ is known as effective potential. 

The conditions for the light to move along the circular orbit are $\dot{r}=0$ and $\ddot{r}=0$, which can be translated to
\begin{eqnarray}
V(r)=\frac{1}{b^2}~~~~~\text{and}~~~~~ V^{'}(r) = 0.
\label{con1}
\end{eqnarray}
Here prime stands for the first derivative with respect to radial coordinare $r$. The circular orbit is called photon
sphere. It is stable or unstable depending on whether the second order derivative of effective potential
is positive or negative. The unstable photon sphere plays a crucial role in the gravitational lensing as we shall see later.

\begin{figure}
\begin{center}
\includegraphics[width=0.8\textwidth]{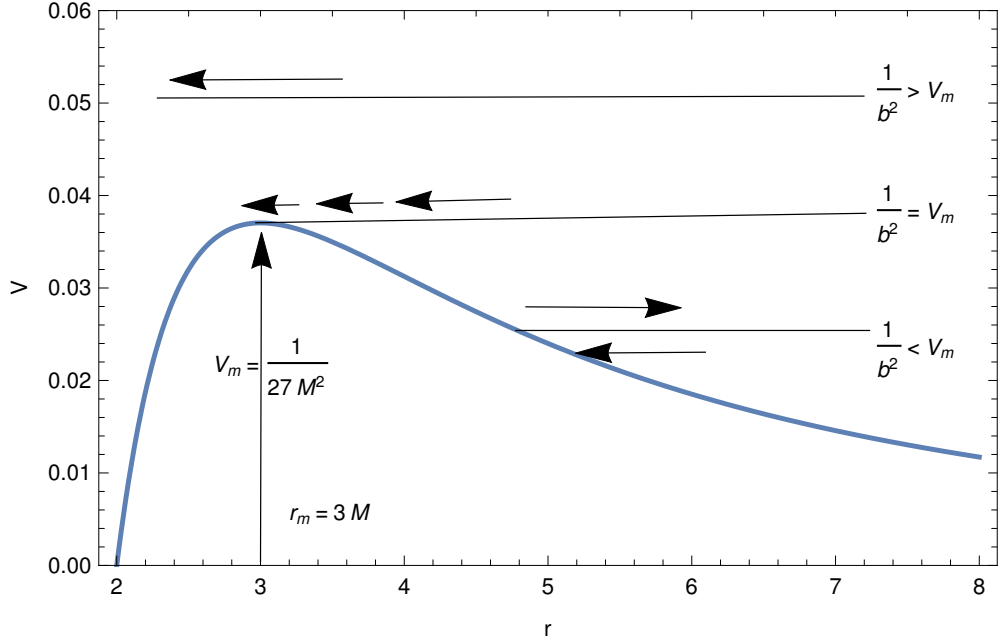}
\caption{\label{vsch}
In this figure we plot the effective potential for radial motion of the light ray traveling on the equatorial plane of Schwarzschild black hole. We have taken M=1 to make this plot. Effective potential starts from zero at the event horizon, it increases and admits maximum at $r=3M$ with the peak value $V_m=\frac{1}{27M^2}$. It is called photon sphere. Then it decreases and goes to zero at infinity. The ingoing light ray with impact parameter $b$ with $\frac{1}{b^2}<V_m$ gets reflected back at the radial location outside the photon sphere. If $\frac{1}{b^2}>V_m$, the light ray enters event horizon.
If $\frac{1}{b^2}=V_m$ light ray asymptotically approaches the photon sphere.
}
\end{center}
\end{figure}

Now we now focus our attention on Schwarzschild spacetime. The metric is given by the expression
\begin{equation}
 ds^2=-\left(1-\frac{2M}{r}\right)dt^2+\frac{dr^2}{\left(1-\frac{2M}{r}\right)}+r^2\left(d\theta^2+\sin^2{\theta}~ d\phi^2 \right)
 \label{sch}.
\end{equation}
It represents a black hole whose event horizon is located at $r=2M$.
Non-vanishing components of four-velocity of the light ray traveling along the equatorial plane of Schwarzschild black hole 
can be obtained by substituting  $f(r)=\left(1-\frac{2M}{r} \right)$ in Eq.\ref{eq2} and are given by 
\begin{eqnarray}
  &&\dot{t}=\frac{1}{b \left(1-\frac{2M}{r} \right)} \nonumber \\
  &&\dot{\phi}=\pm \frac{1}{r^2} \nonumber \\
  &&\dot{r}^2+\frac{1}{r^2}\left(1-\frac{2M}{r} \right)= \frac{1}{b^2}. 
 \label{eq3}
\end{eqnarray}
The equation for the radial motion can be rewritten using Eq.\ref{ve} and is given by 
\begin{equation}
 \dot{r}^2+V(r)=\frac{1}{b^2}~~~~~\text{with}~~~~~ V(r)=\frac{1}{r^2}\left(1-\frac{2M}{r} \right),
 \label{ve1}
\end{equation}
The location of photon sphere can be obtained easily by solving Eq.\ref{con1}. The photon sphere is situated at the radial location 
\begin{equation}
 r_{ph}=3M. 
 \label{phs}
\end{equation}
It is an unstable photon sphere. The maximum value of the potential and impact parameter associated with it is given by
\begin{equation}
 V_m=V(r_{ph})=\frac{1}{27M^2}~~~~\text{and}~~~~b_{ph}=3\sqrt{3}M
 \label{max}
\end{equation}
We note that the height at the peak is inversely proportional to the square of the mass while impact parameter is directly proportional 
to the mass of the black hole, which will be used later.

The effective potential is depicted in Fig.1. It is zero at the event horizon. It increases as we move outwards. It admits a maximum at the photon sphere
located at $r=3M$. Then it goes on decreasing and goes to zero at infinity which is the manifestation of the fact that Schwarzschild black hole spacetime is asymptotically flat.

Consider a light ray moving in the radially inward direction starting from infinity. From Eq.\ref{ve1} it is evident that motion of light ray in the radial direction is equivalent to the motion of particle moving under a potential in one dimension. If $\frac{1}{b^2}<\frac{1}{27M^2}$, the light will encounter the potential barrier and it will be reflected back in the outward direction at a radial location outside the photon sphere. If $\frac{1}{b^2}>\frac{1}{27M^2}$ then the light will continue moving in the inward direction as it does not encounter potential barrier and it would enter the event horizon of the black hole. If $\frac{1}{b^2}=\frac{1}{27M^2}$ it will asymptotically approach the photon sphere as both $\dot{r}$ and $\ddot{r}$ would tend to zero close to photon sphere. However since the angular velocity 
is non-zero it will revolve around the black hole infinitely many times. We will be interested in the case where $\frac{1}{b^2}$ is slightly less that $\frac{1}{27M^2}$. In this case light ray will eventually turn back, but it will spend a lot of time close to the photon sphere. Since angular velocity is finite, it will revolve around the black hole large number of times and thus suffering from large deflection, which is the situation of interest to us.  

\section{Gravitational Lensing Formalism}

In this section we review the basic gravitational lensing formalism employed in this paper to compute location of images, Einstein rings and their properties. Gravitational lensing calculation has two basic components, lens diagram and deflection angle. Lens diagram allows us to relate the location of image to the location of source given the total deflection suffered by the light as it travels from source to observer. So we need to compute the deflection angle which is the only input from general relativity that can be obtained by integrating the geodesic equations. 

We assume that the source is almost exactly behind the black hole so that source, black hole and observer are nearly aligned. We also assume that both source and observer are sufficiently far from the length scale over which spacetime acting as a lens is curved, which is approximately size of the event horizon of the black hole. Thus for all practical purposed lens can be thought of as a point. Except for in the vicinity of the black hole where light suffers deflection, light would travel in the straight line. Thus we can assume that the we are dealing with flat space with one point in the middle acting as a lens. Light travels in the straight line and only when it encounters the lens it suffers from the deflection as depicted in the lens diagram Fig.2.

\begin{figure}
\begin{center}
\includegraphics[width=0.6\textwidth]{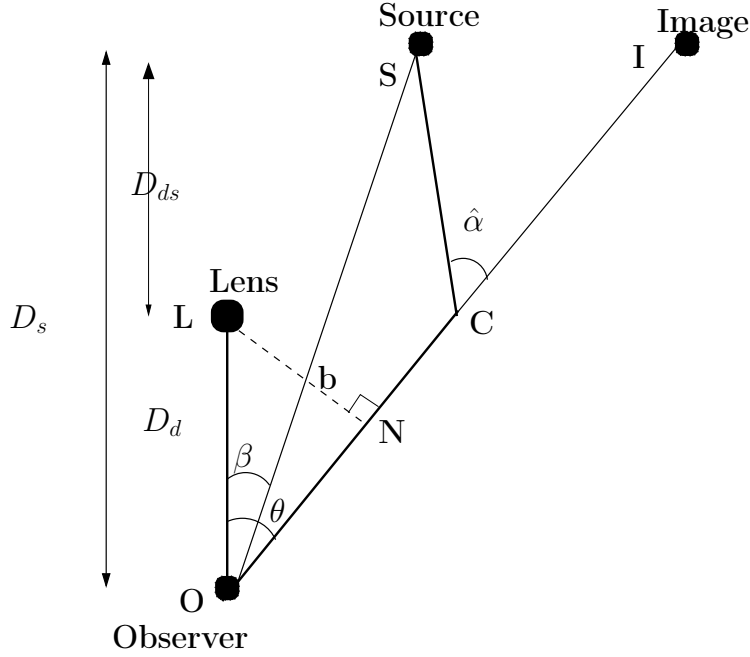}
\caption{\label{Lens Diagram}
Lens diagram is depicted in this picture. L, S, I, O stand for the lens, source, image and observer. $D_d$, $D_{ds}$, $D_s$ stand for the distances between lens and observer, lens and source, source and observer respectively. Angles $\beta$ and $\theta$ stand for the source and image location. Angle $\hat{\alpha}$ stands for the deflection angle. $b$ is am impact parameter. 
}
\end{center}
\end{figure}

In the lens diagram L stands for the lens which is the black hole. S and O stand for the source and observer respectively which are located farway in the asymptotically flat region. The line joining observer and lens is known as optic axis. In the absence of any lens light will travel along the straight line SO. It makes an angle $\beta$ with respect to the optic axis and depicts the source location with respect to the observer. In the presence of lens it will suffer deflection at point C, with the deflection angle $\hat{\alpha}$. It will travel along OC. It makes an angle $\theta$ with respect to the optic axis. Hence the light seems to originate from I which is the image and $\theta $ depicts the image location for observer. $D_d$, $D_{ds}$ and $D_s$ are the distances between observer and lens, lens and source, and observer and source respectively.

Since we are dealing with the near-aligned situation, angles $\beta$ and consequently angle $\theta$ would be very small. Deflection angle $\hat{\alpha}$ can be very large as light ray can go around the lens multiple times. It will be approximately equal to the multiple of $2\pi$. LN is perpendicular to the OI drawn from the lens and b is an impact parameter. From the lens diagram we get 
\begin{equation}
 \sin{\theta}=\frac{b}{D_d}
 \label{sth}
\end{equation}
The location of image can be related to the location of source given the deflection angle $\hat\alpha$ by following relation,
\begin{equation}
 \tan{\beta}=\tan{\theta}-\frac{D_{ds}}{D_s}\left( \tan{\theta}+\tan{\left( \hat{\alpha}-\theta \right)} \right).
 \label{le}
\end{equation}
The equation above is known as Virbhadra-Ellis lens equation which is very popular in the literature. There are other equations available in literature which include the exact lens equation \cite{ns1},\cite{els}. The comparison of various lens equations appears in \cite{lscom}. As quoted there Virbhadra-Ellis lens equation is put in the very simple form that is very easy to use in various realistic situations as it is expressed in terms of distance between source, lens and observer planes.

The light ray emitted by the source will travel in inward direction towards the black hole. Since we are interested in gravitational lensing, we focus on those light rays which get reflected back at some radial location $r_0$ outside the photon sphere and reach observer. The total amount of deflection suffered by the light in its journey from source to observer is given by the following expression,
\begin{eqnarray}
 \hat{\alpha}(r_0)&=& 2 \int_{r_0}^{\infty}\frac{d\phi}{dr} dr-\pi 
 =2 \int_{r_0}^{\infty}\frac{\dot{\phi}}{\dot{r}} dr-\pi  \nonumber \\
                  &=&  2 \int_{r_0}^{\infty} \frac{dr}{r^2\sqrt{V(r_0)-V{(r)}}} -\pi.
 \label{alph}
\end{eqnarray}
To derive the equation above we have used Eq.\ref{eq2} and Eq.\ref{ve}. We have also used the result
\begin{equation}
 b=\frac{1}{\sqrt{V(r_0)}},
 \label{vb}
\end{equation}
which follows from Eq.\ref{ve}.

The way we obtain images from the equations given above in this section is as follows. We obtain deflection angle $\hat{\alpha}$ as a function of turning point $r_0$ using Eq.\ref{alph}. Using Eq.\ref{sth} and Eq.\ref{vb} we express it as a function of $\theta$. Then we substitute that into the lens equation. i.e. Eq.\ref{le}. For a given value of $\beta$ we solve equation to obtain 
the values of $\theta$, i.e. image locations. For non-zero value of $\beta$ we can get multiple images. When $\beta=0$ we get Einstein rings since light which starts off from the source can reach observer from all possible directions and hence the solution to lens equation gives us the radii of Einstein rings. Solving lens equation is a daunting task because it is a complicated transcendental equation. But under the assumptions we make it can be solved analytically \cite{bozza}. We had developed an approach based on effective potential which we employ here to obtain images \cite{bbh}. 

The magnification of images $\mu$ is given by 
\begin{equation}
\mu=\frac{\sin{\theta}}{\sin{\beta}}\frac{d\theta}{d\beta}=\mu_t \mu_r,
\label{mag1}
\end{equation}
where $\mu_t$ and $\mu_{r}$ are tangential and radial magnifications respectively with $\mu_t=\frac{\sin{\theta}}{\sin{\beta}}$ and $\mu_r=\frac{d\theta}{d\beta}$. The divergence of tangential and radial magnifications are called tangential caustic and radial caustic. Quite evidently Einstein rings formed when $\beta=0$ correspond to the tangential caustic.

Since $\beta$ and $\theta$ are very small and deflection angle $\hat{\alpha}$
is approximately multiple of $2\pi$, Eq.\ref{sth} and Eq.\ref{le} can be approximated to 
\begin{eqnarray}
 &&\theta=\frac{b}{D_d} \nonumber \\
 &&\beta=\theta -\frac{D_{ds}}{D_s}\delta\alpha_n \nonumber \\
 &&\alpha=2\pi n + \delta\alpha_n ~~~~~\text{with}~~~~~|\delta\alpha_n|<<1
 \label{appeq}
\end{eqnarray}
We use effective potential formalism to write approximate expression for $\delta\alpha_n$ and then use the equations above to obtain image locations. Further the value of tangential magnification can be approximated to 
\begin{equation}
 \mu_{t}= \frac{\theta}{\beta}.
 \label{appmag}
\end{equation}

Different light rays that circle the black hole different number of times follow different trajectories in space and thus would take different amount of time to reach the observer although they might originate from the source at the same time. The time taken by time to depart from the source and reach the observer is given by 
\begin{eqnarray}
 t&=&2\int_{r_0}^{D_d}\frac{d\phi}{dr}dr =2\int_{r_0}^{D_d}\frac{\dot{\phi}}{\dot{r}}dr  \nonumber \\
 &=&2\int_{r_0}^{D_d}\frac{\sqrt{V(r_0)}}{f(r)\sqrt{V(r_0)-V(r)}} dr,
\label{tm}
\end{eqnarray}
Where $D_d$ is the distance between lens and source or observer which is very large, thus if needed it can be taken to be infinity when it would be appropriate to do so. Since observer and source are located at large distance from the lens, light will take large amount of time to reach lens and it will diverge if we set upper limit of integration to infinity. Thus we have $D_d$ as the upper limit of integration and not infinity as we had while calculating deflection angle. The light will also spend a lot of time circling the black hole depending on number of times. We compute the time delay between arrival of light rays which circle around the black hole different number of times.

We conclude the discussion on gravitational lensing formalism here. In the next section we describe the gravitating system of black hole surrounded by matter that we consider in this paper.

\section{Black hole with matter around it}

We now describe the situation under consideration as depicted in Fig.3, where we have a black hole with significant amount of matter surrounding it. The mass associated with black hole is $M_1$. So its event horizon is located at $r=2M_1$ and photon sphere is located at $r= 3M_1$. The mass outside the matter distribution is $M_2$, so the photon sphere associated with it is located at $r=3M_2$. Further the height of the potential at the inner photon sphere where it admits a peak is higher than the height of the peak at outer photon sphere, since potential at maximum scales inversely with the square of the mass. We assume that matter is located from radial location $r_1$ to $r_2$ which lie between the two photon spheres. So that we have $3M_1<r_1<r_2<3M_2$. 

\begin{figure}
\begin{center}
\includegraphics[width=1.0
\textwidth]{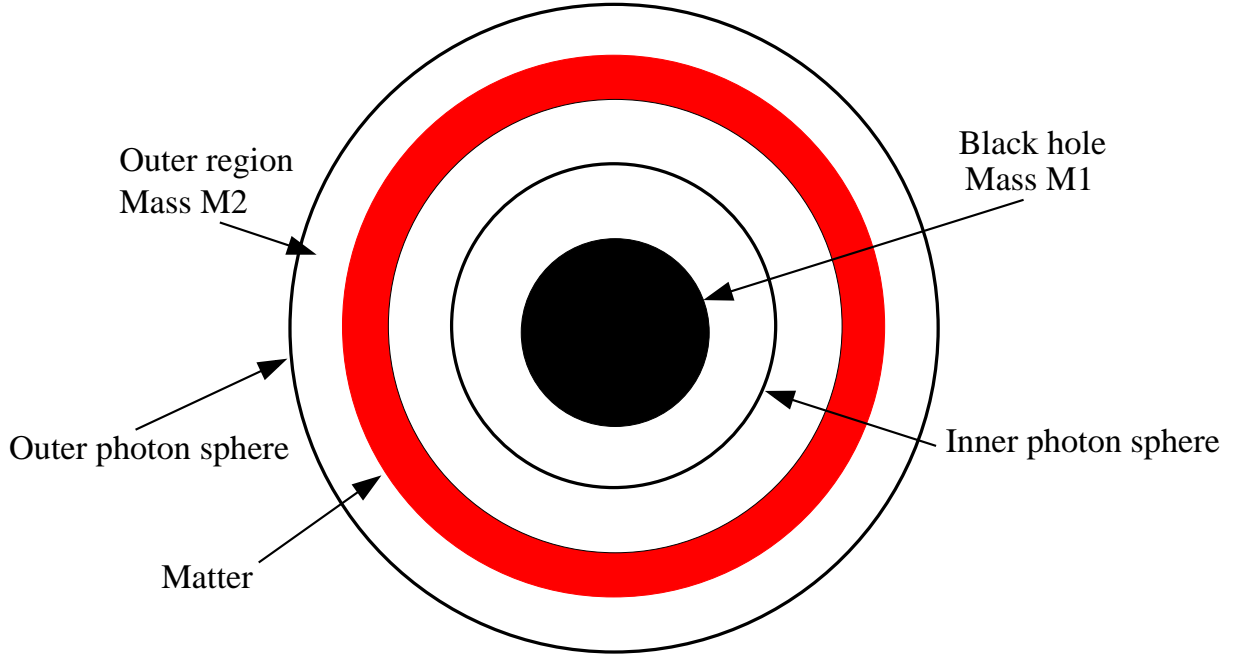}
\caption{\label{matd}
In this figure we depict the gravitating system acting as a lens under investigation. We have a black hole with mass $M_1$ at the center. The photon sphere associated with it i.e. inner photon sphere is depicted in the figure. The mass associated with outer region is $M_2$ and the photon sphere associated with it i.e. outer photon sphere is depicted in the figure. Matter is distributed in the region between two photon spheres.
}
\end{center}
\end{figure}

We assume that the metric inside the region where matter is located is given by 
\begin{equation}
 ds^2=-\left(1-\frac{2m(r)}{r}\right)dt^2+\frac{dr^2}{\left(1-\frac{2m(r)}{r}\right)}+r^2\left(d\theta^2+\sin{\theta}^2 d\phi^2\right),
\end{equation}
where $m(r)$ is monotonically increasing function which interpolates between 
$M_1$ at $r=r_1$ to $M_2$ at $r=r_2$. The effective potential is given by the expression 
\begin{equation}
 V(r)=\frac{1}{r^2}\left(1-\frac{2m(r)}{r} \right),
 \label{ef3}
\end{equation}
which we assume to be the monotonically decreasing function in the region where matter is located. This is true if $\left( r-3m(r)+r m^{'}(r) \right)>0$. So effective potential is continuous at $r=r_1$ and $r=r_2$ and also there is no unstable photon sphere in this region. 

The overall the effective potential looks as depicted in the Fig.4. It admits the value zero at the event horizon of the black hole. It goes on increasing and admits a maximum with the value at peak $V=\frac{1}{27 M_1^2}$ at the location of inner photon sphere located at $r=3M_1$. It then decreases  until the inner boundary of matter distribution $r=r_1$. It further decreases within the matter distribution all the way upto outer edge at $r=r_2$. It then increases in the outer region and admits maximum at $r=3M_2$ with the peak value $V=\frac{1}{27 M_2^2}$. The outer peak has less height as compared to the inner peak. It then decreases and goes to zero at infinity. 

\begin{figure}
\begin{center}
\includegraphics[width=0.85\textwidth]{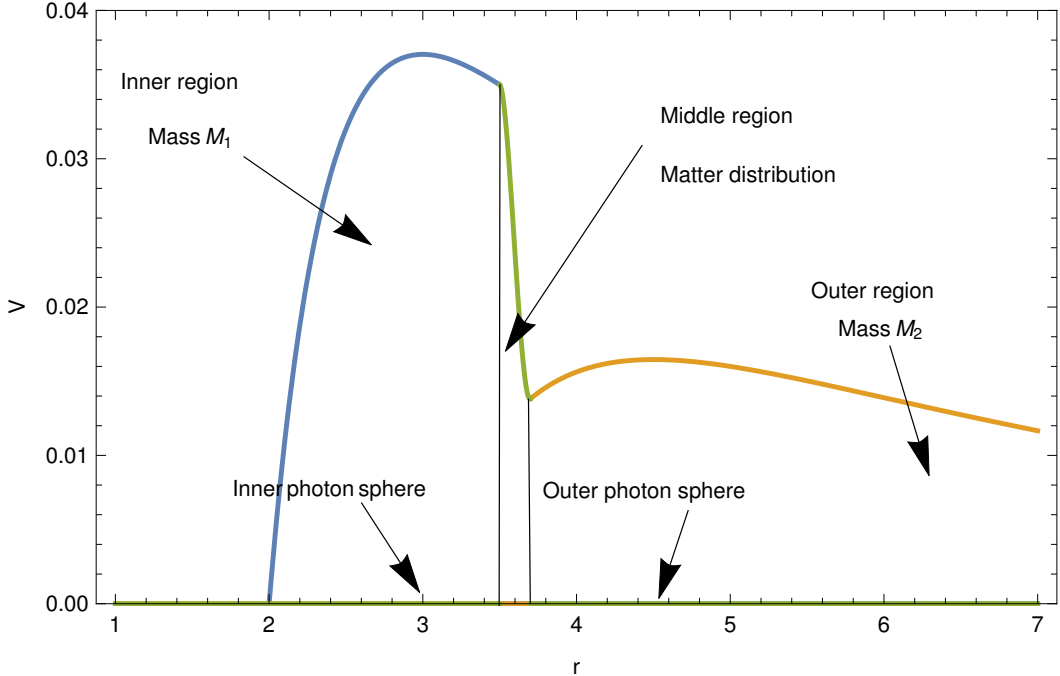}
\caption{\label{ef2}
In this figure we plot the effective potential for radial motion. The mass associated with inner region is $M_1$ and outer region is $M_2$. In order to make this plot we have chosen $M_1=1$ and $M_2=1.5$. Locations of inner photon sphere and outer photon sphere are depicted in the figure. The peak value at inner photon sphere is larger than that at outer photon sphere. Matter distribution extends between the region in between two photon spheres. The effective potential is a monotonically decreasing function inside the matter region. 
}
\end{center}
\end{figure}

This is very peculiar behaviour which leads to novel features in the distribution of images when we study the gravitational lensing. We get three distinct set of infinitely many images as opposed to just one set for isolated Schwarzschild black hole. We get one set of images due to the light rays that get reflected back close to the outer photon sphere. We get second set of images due to the light rays that enter the outer photon sphere just above the peak and get reflected back in the region where matter is situated.
We get third set of images due to the light rays that get reflected back just outside the inner photon sphere. The region where is matter is located will be denoted with subscript $m$. The inner region and outer regions will be denoted by the subscript $1$ and $2$ respectively throughout the manuscript.

\section{Images due to the light rays that turn back outside outer photon sphere}
In the section we discuss the ingoing light rays from the source for which $\frac{1}{b^2}$ is slightly less that the maximum $V_{2m}=\frac{1}{27M_2^2}$ which get reflected back off the potential just outside the outer photon sphere located at $r_{ph2}=3M_2$. Since the effective potential admits maximum at the photon sphere, in its vicinity both $\dot{r}$ and $\ddot{r}$ are very small and thus light ray spends a lot of time circling around the black hole with finite angular speed. Thus deflection angle suffered by it is very large. Light can go around the black hole once, twice and so on all the way upto infinitely many turns which results into formation of infinitely many images. 

Range of radial coordinate $r$ is infinite since it varies from $r_0$ which is slightly above radial location of outer photon sphere $r_{ph2}$ to infinity. We redefine radial coordinate in order to make the range finite. We define coordinate $y$ as 
\begin{equation}
 y=\frac{f_{2}(r)-f_2(r_0)}{1-f_2({r_0})} 
 \label{y2}
\end{equation}
where $f_2(r)=\left(1-\frac{2M_2}{r} \right)$. It is quite evident from the definition above that $y$ is monotonically increasing function of $r$ and it varies from $0$ to $1$ as $r$ varies from $r_0$ to $\infty$. Thus we can invert it and write down radial coordinate $r$ as a function of $y$ as
\begin{equation}
r(y)=\frac{r_0}{1-y}
\label{ry}
\end{equation}
We define a quantity $T_{2}(r_0,r)$ which is the derivative of the two radial coordinates as 
\begin{equation}
 T_2(r_0,r(y))=\frac{dr}{dy}=\frac{1-f_{2}(r_0)}{f_2^{'}(r(y))},
 \label{vc2}
\end{equation}
which will  be useful later.

The deflection suffered by the light as it travels from source to observer is given by 
\begin{equation}
 \hat{\alpha}_2= I_2 -\pi,
 \label{d2}
\end{equation}
where $I_2$ is given by 
\begin{eqnarray}
 I_2 &=& 2\int_{r_0}^{\infty} \frac{1}{r^2\sqrt{V_{2}(r_0)-V_2(r)}}dr \nonumber \\
   &=& \int_{0}^{1} \frac{F_2(r_0,r(y))}{\sqrt{V_{2}(r_0)-V_2(r(y))}}dy,
 \label{d21}
\end{eqnarray}
where function $F_2(r_0,r(y))$ is given by
\begin{equation}
 F_2(r_0,r(y))=\frac{2T_2(r_0,r(y))}{r(y)^2}.
\end{equation}

Since we are integrating over a finite range and integrand if finite everywhere except at the turning point, the divergence of the integral can arise only due to the divergent behavior at $y=0$.  
While $F_2(r_0,r(y))$ is always a well-behaved function in the entire range of $y$, function $\frac{1}{\sqrt{V_{2}(r_0)-V_2(r(y))}}$ diverges at $y=0$.
Taylor-expanding $V_{2}(r_0)-V_2(r(y))$ around $y=0$, we obtain
\begin{equation}
 V_{2}(r_0)-V_2(r(y))=\alpha_2(r_0)y+\beta_2(r_0)y^2+O(y^3),
\end{equation}
where 
\begin{eqnarray}
\alpha_2(r_0) &=& -T_{2}(r_0,r_0)V_2^{'}(r_0) \nonumber\\
\beta_2(r_0) &=& -\frac{1}{2}\left(T_2(r_0,r_0)T_{2}^{'}(r_0,r_0)V_{2}^{'}(r_0)+ 
T^{2}(r_0,r_0) V_{2}^{''}(r_0)\right).
\label{ab}
 \end{eqnarray}
Quite evidently if the turning point $r=r_0$ is far from the outer photon sphere, Taylor-expansion upto first term would suffice and integral and thus the deflection angle is finite.

The situation is different when the turning point is close to the photon sphere when the first derivative term will be vanishingly small and thus we need to retain in the Taylor-expansion terms upto the second order, leading to the logarithmic divergence of the integral. 

\begin{figure}
\begin{center}
\includegraphics[width=0.85\textwidth]{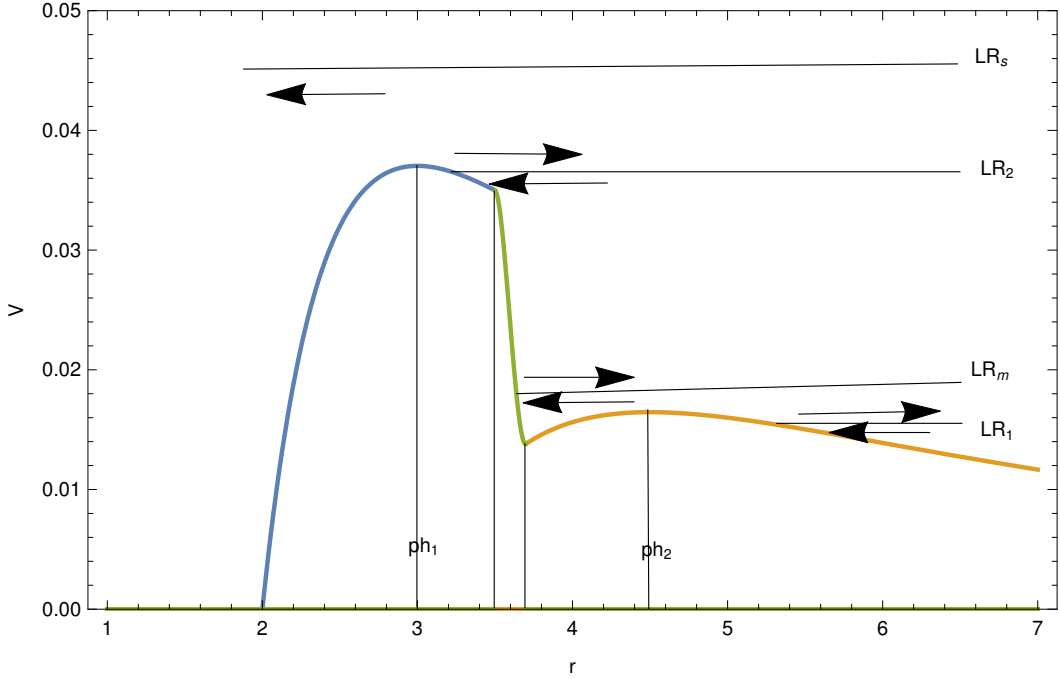}
\caption{\label{phh}
In this figure we depict three different cases of interest to us. First case denoted by $LR_1$ corresponds to the ingoing light rays that admit turning point just outside the outer photon sphere. $LR_m$ corresponds to the light rays that enter the inner photon sphere just above the peak enter into the region where matter is located and turn back. $LR_2$ corresponds to the light rays that turn back in the vicinity of inner photon sphere. $LR_s$ correspond to the light rays that enter the inner photon sphere and are eventually engulfed by the black hole.
}
\end{center}
\end{figure}

To isolate the divergent term and to understand the logarithmic nature of divergence we Taylor expand $\alpha_2(r_0)$ and $\beta_2(r_0)$ around $r_0=r_{ph2}$. We get
\begin{eqnarray}
 \alpha_2(r_0)&=&-T_2(r_{ph2},r_{ph2})V_{2}^{''}(r_{ph2})r_{ph2}\left(1-\frac{r_{ph2}}{r_{0}}\right) +O\left(\left(1-\frac{r_{ph2}}{r_0}\right)^2\right) \nonumber \\
 \beta_2(r_0)&=&-\frac{1}{2}T_2^2(r_{ph2},r_{ph2})V_2^{''}(r_{ph2})+O\left(1-\frac{r_{ph2}}{r_0}\right).
\end{eqnarray}
The divergent piece in the deflection angle is given by 
\begin{eqnarray}
 I_{D2}&=&F_2(r_{ph2},r_{ph2})\int_{0}^{1}\frac{1}{\sqrt{\alpha_{2}(r_0)y+\beta_{2}(r_0)y^2}} dy\nonumber \\
 &=& -A_2\log{\left(1-\frac{r_{ph2}}{r_0}\right)} +\tilde{B}_2+O\left(1-\frac{r_{ph2}}{r_0}\right),
\end{eqnarray}
where 
\begin{eqnarray}
 A_2&=&\frac{F_2(r_{ph2},r_{ph2})}{\sqrt{\beta_2(r_{ph2})}} \nonumber \\
 \tilde{B}_{2}&=& \frac{F_2(r_{ph2},r_{ph2})}{\sqrt{\beta_2(r_{ph2})}} \log{\left( \frac{2T_{2}(r_{ph2},r_{ph_2})}{r_{ph2}}   \right)},
\end{eqnarray}
and regular piece is given by 
\begin{eqnarray}
 I_{2R}&=&\int_{0}^{1}\left( \frac{F_2(r_0,r(y))}{\sqrt{V_{2}(r_0)-V_2(r(y))}} - \frac{F_2(r_{ph2},r_{ph2})}{\sqrt{ \alpha_2(r_0)y +  \beta_{2}(r_0)y^2  }}\right)dy \nonumber \\ \\
 &=& I_{2R}(r_{ph2}) + O\left(1-\frac{r_{ph2}}{r_0}\right).
\end{eqnarray}

Combining divergent and convergent pieces we can write down deflection angle as 
\begin{equation}
 \hat{\alpha}_2=-A_2 \log{\left(B_2\left(1-\frac{r_{ph2}}{r_0}\right)\right)}-\pi +  O\left(1-\frac{r_{ph2}}{r_0}\right),
 \label{deff2}
\end{equation}
where $B_2$ is given by 
\begin{equation}
 B_2=exp\left(-\frac{\tilde{B}_2+I_{2R}(r_{ph2})}{A_2}\right).
\end{equation}
Thus we have demonstrated the logarithmic divergence of deflection angle we had anticipated earlier.

We can also Taylor-expand the impact parameter around the outer photon sphere. Using Eq.\ref{vb} we get
\begin{equation}
 b_2=C_2+D_2\left(1-\frac{r_{ph2}}{r_0}\right)^2,
 \label{b2f}
\end{equation}
where 
\begin{eqnarray}
 C_2 &=&\frac{1}{\sqrt{V_{2}(r_{ph2})}} \nonumber \\
 D_2 &=&-\frac{1}{4} \frac{ V_{2}^{''}(r_{ph2})}{V_2^{\frac{3}{2}}(r_{ph2})} r_{ph2}^2.
\end{eqnarray}

Using Eq.\ref{appeq}, Eq.\ref{deff2}, Eq.\ref{b2f} we get the location as images as 
\begin{equation}
 \theta_{2,n}=\frac{C_2}{D_d}+\frac{1}{D_d}\frac{D_2}{B_2^2} exp\left( -\frac{2(2n+1)\pi}{A_2}\right) \left(1+\frac{2}{A_2}\frac{D_s}{D_{ds}}\beta \right),
 \label{im2}
\end{equation}
where the subscript $n$ stands for the number of times light winds around the black hole in its journey from source to observer. Since the deflection angle diverges as the turning point approaches the outer photon sphere, all values of $n$ all the way upto infinity are realized leading to the formation of infinitely many relativistic images. All the images lie on right side of critical angle 
\begin{equation}
 \bar{\theta}_2=\frac{C_2}{D_d}=\frac{3\sqrt{3}M_2}{D_d},
 \label{cr1}
\end{equation}
and separation from the critical angle goes on decreasing exponentially as winding number $n$ increases.

From Eq.\ref{mag1}, Eq.\ref{im2} we obtain the expression for the magnification of images as
\begin{equation}
 \mu_{2,n}=\frac{2}{\beta}\frac{D_s}{D_{ds}D_d^2}\frac{D_2}{A_2 B_2^2}\left(\frac{D_2}{B_2^2}exp\left( -\frac{4(2n+1)\pi}{A_2} \right) + C_2 exp\left( -\frac{2(2n+1)\pi}{A_2} \right)\right).
\end{equation}

These are the results for the clockwise winding of light around black hole when images lie on the same side as that of source. For the counter-clockwise winding the location of images are given by 
\begin{equation}
 \bar{\theta}_{2,n}=-\frac{C_2}{D_d}+\frac{1}{D_d}\frac{D_2}{B_2^2} exp\left( -\frac{2(2n+1)\pi}{A_2}\right) \left(-1+\frac{2}{A_2}\frac{D_s}{D_{ds}}\beta \right),
 \label{im2c}
\end{equation}
Magnification diverges when $\beta=0$ i.e. when the source is exactly behind the lens which corresponds to the tangential caustic. Since light emitted by source can reach observer from all possible directions we get rings as images, known as Einstein rings. The radii of Einstein rings are given by 
\begin{equation}
 \theta_{2E,n}=\frac{C_2}{D_d}+\frac{1}{D_d} \frac{D_2}{B_2^2} Exp\left( -\frac{2(2n+1)\pi}{A_2}\right).
\end{equation}
As it can be seen from the expression above, all rings lie beyond the critical angular radius $\bar{\theta}_2$. There are infinitely many rings with separation of rings from critical value decreasing as $ n $ increases.

We now compute the time required for the light ray to originate from the source and reach observer. The time is given by the expression
\begin{eqnarray}
 t_2 &=& 2\int_{r_0}^{D_d}\frac{\sqrt{V_2(r_0)}}{f_2{(r)}}\frac{1}{\sqrt{V_2(r_0)-V_2(r)}}dr \nonumber \\
 &=& 2\int_{0}^{y_{max}}\frac{\sqrt{V_2(r_0)}}{f_2{(r(y))}}T_2(r_0,y)
 \frac{1}{\sqrt{V_2(r_0)-V_2(r(y))}}dy \nonumber \\
 &=& 2\int_{0}^{y_{max}} G_{2}(r_0,y) g_{2}(r_0,y)dy.
 \label{tm1}
\end{eqnarray}
Here $y_{max}$ corresponds to the location of source and observer from lens, which is a number very close to one. $G_{2}(r_0,y)$ and $g_2(r_o,y)$ are given by 
\begin{eqnarray}
 G_{2}(r_0,y)&=&\frac{\sqrt{V_2(r_0)}}{f_2{(r(y))}}T_2(r_0,y), \nonumber \\
 g_{2}(r_0,y)&=&  \frac{1}{\sqrt{V_2(r_0)-V_2(r(y))}}.\label{tdd}
\end{eqnarray}

There are two reason this integral will diverge. First because integrand in the first line of Eqn.\ref{tm1} is finite for the large values if $r$ and we integrate upto extremely large value of $r$. So the divergence is sourced by infinity. Secondly because integrand will diverge due to the factor of $(V_2(r_0)-V_2(r))$ in the denominator at $r_0$ which is very close to the outer photon sphere located at $r=r_{ph2}$. 

We now write down the term which captures the divergence sourced from infinity as
\begin{equation}
 I_{D2,1}=2\int_{0}^{y_{max}} G_2({r_{ph2}},y)~g_2({r_{ph2}},y=y_{max})dy,
 \label{td1}
\end{equation}
which can be evaluated from Eq.\ref{ry}, Eq.\ref{vc2}, Eq.\ref{tdd}. and turns out to be 
\begin{equation}
 I_{D2,1}=\frac{4M_2}{(1-f_2(r_{ph2}))} \int_{0}^{y_{max}}
 \frac{1}{(1-y)^2 ( f_2(r_{ph2})+(1-f_2(r_{ph2}))y )}~dy.
\end{equation}
Upon computing the integral we get 
\begin{equation}
  I_{D2,1}=2D_d -4M_2\log{\left(\frac{3M_2}{D_d}\right)}-6M_2-4M_2\log{(f_2(r_{ph2}))}
  \label{id21}.
\end{equation}
The first two terms in the expression above are extremely large because of the largeness of distance $D_d$. The rest of the terms are finite.

We now try to capture the divergence arising because of the proximity of
turning point to the outer photon sphere. As before upon Taylor expanding $(V_2(r_0)-V_2(r))$ and keeping the terms upto second order we get the divergent term $I_{D2,2}$.
\begin{equation}
 I_{D2,2}= 2 \int_{0}^{1} G_2(r_{ph2},y=0)\frac{1}{\sqrt{\alpha_2 y +\beta_2 y^2}} dy,
\end{equation}
where $\alpha_2$, $\beta_2$ are the terms as defined earlier in Eq.\ref{ab}. Upon computing the integral and Taylor expanding around the outer photon sphere we get 
\begin{equation}
 I_{D2,2}=-\bar{A_2}\log{\left(1-\frac{r_{ph2}}{r_0}\right)}+\bar{B}_2+O\left(1-\frac{r_{ph2}}{r_0}\right),
\end{equation}
where $\bar{A_2}$ and $\bar{B_2}$ are given by
\begin{eqnarray}
 \bar{A}_2&=&\frac{2G_2({r_{ph2}},y=0)}{\sqrt{\beta_2(r_{ph2})}} \nonumber \\
  \bar{B}_2&=&\frac{2G_2({r_{ph2}},y=0)}{\sqrt{\beta_2(r_{ph2})}} \log{\left(\frac{2T_2(r_{ph2},r_{ph2})}{r_{ph2}} \right)}.
\end{eqnarray}

The regular part of the integral is given by
\begin{eqnarray}
 I_{R2,2}(r_0)&=& 2\int_{0}^{y_{max}} G_{2}(r_0,y) g_{2}(r_0,y)dy ~-
 2\int_{0}^{y_{max}} G_{2}(r_{ph2},y) g_{2}(r_{ph2},y=1)dy \nonumber \\
 &&-2\int_{0}^{1} G_{2}(r_{ph2},y=0)\frac{1}{\sqrt{\alpha_2 y +\beta_2 y^2}} dy \nonumber \\
 &=& I_{R2,2}(r_{ph2})+O\left( 1-\frac{r_{ph2}}{r_0} \right)
 \end{eqnarray}
 
 Combining all the contributions we get an expression for time $t_2$
 \begin{eqnarray}
  t_2(r_0)&=& 2 D_s -4M_2\log{\left(\frac{3M_2}{D_s}\right)} - 
  \bar{A_2}\log{\left( 1-\frac{r_{ph2}}{r_0} \right)}+\bar{B}_2 \nonumber \\
  &&- 6M_2 -4M_2\log{\left(f_2(r_{ph2})\right)} + I_{R2,2}(r_0).
 \end{eqnarray}
 
 Note that not all values of $r_0$ correspond to images. Only those values for which deflection angle is approximately equal to integral multiple of $\pi$ correspond to images. From Eq.\ref{deff2} we gets
 \begin{equation}
  \hat{\alpha}_n=-A_2\log{\left(1-\frac{r_{ph2}}{r_0}\right)}+\tilde{B_2}+I_{2R}(r_{ph2})-\pi \approx 2n\pi.
 \end{equation}
 Combining the two equations above we get time required for light to travel from source to observer leading to the formation of $n^{th}$ image.
 \begin{equation}
  t_{2,n}=2D_d-4M_2\log{\left( \frac{3M_2}{D_d} \right)} +\frac{\bar{A_2}}{A_2}(2n+1)\pi+I_{2R}^{'},
 \end{equation}
where $I_{2R}^{'}$ is given by 
\begin{equation}
 I_{2R}^{'}=-\frac{\bar{A}_2}{A_2} B_2+\bar{B}_2-6M_2-4M_2\log{f_2(r_{ph2})}+\frac{\bar{A}_2}{A_2}I_{2R}(r_{ph2}).
\end{equation}
Thus the time delay between $n^{th}$ order and $m^{th}$ order images is given by 
\begin{equation}
  t_{2,n}- t_{2,m}=2\frac{\bar{A_2}}{A_2}(n-m)\pi.
\end{equation}
This is the time delay between two images that correspond to the same set, i.e. images formed due to the light rays that get reflected outside the outer photon sphere.

\section{Images due to the light rays that enter outer photon sphere just above peak and get reflected inside the matter}

In this section we deal with images formed due to the ingoing light rays which enter the outer photon sphere just above the peak in the effective potential and enter the region where matter is located where they get reflected back. Since the $\frac{1}{b^2}$ is just above the peak in potential at outer photon sphere $V_2(r_{ph2})$ both $\dot{r}$ and $\ddot{r}$ go to zero from above at the location photon sphere. The light ray spends alot of time near photon sphere $r=r_{ph2}$, while it moves along the azimuthal direction with finite angular velocity. Thus it suffers a large deflection. 
As $\frac{1}{b^2} \rightarrow V_2(r_{ph2})$, the deflection shows divergence.
Thus we can have infinitely many images due to the light that winds once, twice and so on, all the way upto infinity.

We define $r_{m}$ as the radial location where the potential inside matter region is equal to the peak of the potential at outer photon sphere,
\begin{equation}
 V_{m}(r_m)=V_2(r_{ph2})=\frac{1}{27 M_2^2}.
\end{equation}
The deflection suffered by the light ray in this case is given by 
\begin{equation}
 \hat\alpha_m=I_m-\pi
\end{equation}
where $I_m$ is given by 
\begin{eqnarray}
I_m&=&2 \int_{r_0}^{r_2}\frac{1}{r^2\sqrt{V_m(r_0)-V_m(r)}}dr+ 2 \int_{r_2}^{\infty}\frac{1}{r^2\sqrt{V_m(r_0)-V_2(r)}}dr  
 \label{def3}
\end{eqnarray}
where $r_0$ is a turning point which lies in the region containing matter.
The first and second integrals correspond to the contribution to the deflection angle from the matter region and outside region respectively.
The first integral is evaluated over the finite range of radial coordinate, while in the second integral, value of radial coordinate varies from $r_2$ to infinity. In order to make the range finite we introduce a new radial coordinate $x$ as 
\begin{equation}
 x=\frac{f_2(r)-f_2(r_2)}{1-f_2(r_2)}.
\end{equation}
$x$ is a monotonically increasing function of $r$ and it varies from $0$ to $1$ as we vary $r$ from $r_2$ to $\infty$. We can invert this relation and write $x$ in terms of $r$ as
\begin{equation}
 r=\frac{r_2}{1-x}
 \label{rx}
\end{equation}
We define function $T_m$ which is the derivative of two variables as
\begin{equation}
T_m(r(x))=\frac{dr}{dx}=\frac{1-f_2(r_2)}{f_2^{'}(r(x))}.
\label{tmm}
\end{equation}
Thus the integral can be written as 
\begin{equation}
 I_{m}=2 \int_{r_0}^{r_2}\frac{1}{r^2\sqrt{V_m(r_0)-V_m(r)}}dr+  \int_{0}^{1}\frac{F_m(r(x))}{\sqrt{V_m(r_0)-V_2(r(x))}}dx,
 \label{def4}
\end{equation}
where $F_m(r(x))=\frac{2}{r(x)^2}T_m(r(x))$.

The first integral in the Eq.\ref{def4} is computed over the finite range of radial coordinate. The integrand is finite everywhere except for at turning point $r_0$. But since close to the turning point we have 
\begin{equation}
 V_{m}(r_0)-V_m(r)=\alpha_{3}(r_0)(r_0-r),
\end{equation}
where $\alpha_{3}(r_0)=-\frac{dV_{m}}{dr}(r_0)$ is a finite non-zero number, integral turns out to be finite. Thus first term in the integral is finite.

The divergence can however arise due to the second term if $\frac{1}{b_2^2}=V_m(r_0)$ is close to $V_{m}(r_m)=V_{2}(r_{ph2})$, i.e. if turning point $r_0$ is close to the $r_m$. We integrate over finite range and the term in the numerator i.e. $F_m(r(x))$ is finite everywhere. Thus the divergence can come from the term in the denominator $V_m(r_0)-V_2(r(x))$ which can go to zero as the light ray travels just above the peak of potential at the location of outer photon sphere.

We Taylor-expand $V_m(r_0)-V_2(r(x))$ around $x=x_{ph2}$, where $x_{ph2}$ is the location of outer photon sphere expressed in terms of $x$ coordinate. We get
\begin{equation}
 V_{m}(r_0)-V_2(r(x))=\alpha_m(r_0)+\beta_m(x-x_{ph2})^2 +O\left( (x-x_{ph2})^3 \right),
\end{equation}
where 
\begin{eqnarray}
 \alpha_m(r_0)&=&V_{m}(r_0)-V_2(r_{ph2})=V_{m}(r_0)-V_{m}(r_m) \nonumber \\
 \beta_m&=& -\frac{1}{2}T_m^2(r_{ph2})V_{2}^{''}(r_{ph2}).
 \label{alm}
\end{eqnarray}

If $r_0$ is sufficiently away from $r_m$, $\alpha(r_0)$ is finite. Thus Taylor-expansion upto the leading constant term would suffice. Integrand is finite at the location of outer photon sphere and consequently the deflection angle is finite. However if $r_0$ is close to $r_m$ the leading term approaches zero and we have to Taylor-expand upto the second order. The integral would diverge logarithmicaly. If $r_0$ is close to $r_m$ then we can write $\alpha(r_0)$ as 
\begin{equation}
 \alpha_m(r_0)=-V_{m}^{'}(r_{m}) \,r_m\left(1-\frac{r_0}{r_m}\right) + O\left(\left(1-\frac{r_0}{r_m}\right)^2 \right).
\end{equation}
We can isolate the divergent part of the integral as 
\begin{eqnarray}
 I_{Dm}&=&\int_{0}^{1}\frac{F_m(r_{ph2})}{\sqrt{\alpha_m+\beta_m (x-x_{ph2})^2}}dx \nonumber \\
 &=& -A_m \log{\left(1-\frac{r_0}{r_m}\right)}+\tilde{B}_m +O\left( \left( 1-\frac{r_0}{r_m}\right) \right) ,
\end{eqnarray}
where 
\begin{eqnarray}
 A_m&=&\frac{F_{m}(r_{ph2})}{\sqrt{\beta_m(r_{ph2})}} \nonumber \\
 \tilde{B}_{m}&=& \frac{F_{m}(r_{ph2})}{\sqrt{\beta_m(r_{ph2})}}\log{\left( \frac{4 x_{ph2}(1-x_{ph2})\beta_m}{-V_{m}^{'}(r_{m})r_{m}} \right)}.
\end{eqnarray}
The regular part of the integral is given by 
\begin{eqnarray}
 I_{Rm}=&&  2 \int_{r_0}^{r_2}\frac{1}{r^2\sqrt{V_m(r_0)-V_m(r)}}dr \nonumber \\
 &&+  \int_{0}^{1}\left(\frac{F_m(r(x))}{\sqrt{V_m(r_0)-V_2(r(x))}} - 
 \frac{F_m(r_{ph2})}{\sqrt{\alpha_m+\beta_m (x-x_{ph2})^2}} 
 \right)dx \nonumber \\
 =&&  I_{R2}(r_m) + O\left(1-\frac{r_0}{r_m} \right).
\end{eqnarray}

Thus the deflection angle can be written as 
\begin{equation}
 \hat{\alpha}_{m}= -A_m\log{\left( B_{m}\left(1-\frac{r_0}{r_m} \right) \right)}-\pi + O\left(1-\frac{r_0}{r_m} \right),
 \label{defm}
\end{equation}
where 
\begin{equation}
 B_m=Exp\left(-\frac{I_{Rm}(r_m)+\tilde{B}_m}{A_m}  \right)
\end{equation}

Starting from Eq.\ref{vb} we can Taylor-expand impact parameter as 
\begin{equation}
 b_m=C_m-D_m\left(1-\frac{r_0}{r_m} \right),
 \label{bm}
\end{equation}
where
\begin{eqnarray}
 C_m &=& \frac{1}{\sqrt{V_m(r_m)}} \nonumber \\
 D_m &=& -\frac{1}{2} \frac{V_m^{'}(r_m)}{V_m^{\frac{3}{2}}(r_m)} r_m
\end{eqnarray}

From Eq.\ref{appeq}, Eq.\ref{defm}, Eq.\ref{bm} we can we obtain the location of images which is given by the expression
\begin{equation}
 \theta_{m,n}=\frac{C_m}{D_d}-\frac{1}{D_d}\frac{D_m}{B_m}exp\left(-\frac{(2n+1)\pi}{A_m} \right)\left( 1+\frac{1}{A_m}\frac{D_s}{D_{ds}}\beta \right).
 \label{imm}
\end{equation}
Again $n$ is stands for the number of times light ray winds around the black hole. Since the deflection angle shows divergence all values of n are possible all the way upto infinity. It is clear from the expression above all the images lie below the critical angle 
\begin{equation}
 \bar{\theta}_m=\frac{C_m}{D_d}=\frac{3\sqrt{3}M_2}{D_d},
 \label{cr2}
\end{equation}
and the angular separation between the images from $ \bar{\theta}_m$ goes on decreasing as we increase value of $n$. It is clear from Eq.\ref{cr1}, Eq.\ref{cr2} that two critical angles are equal.
\begin{equation}
   \bar{\theta}_2=\bar{\theta}_m.
\end{equation}
Thus images due to the light rays that are reflected back slightly below the outer photon sphere and images due to the light rays that enter outer photon sphere just above the peak are cluttered together above and below the critical angle stated above. 

From Eq.\ref{mag1}, Eq.\ref{imm} we get magnification of the images
\begin{equation}
 \mu_{m,n}=\frac{1}{\beta}\frac{D_s}{D_{ds}D_d^2}\frac{D_m}{A_m B_m}\left(\frac{D_m}{B_m}exp\left( -\frac{2(2n+1)\pi}{A_m} \right) - C_m exp\left( -\frac{(2n+1)\pi}{A_m} \right)\right).
\end{equation}

Results obtained so far are for the light rays that go around the black hole in clockwise sense and images which lie on the same side of optic axis as the source. Location of images which lie on the opposite side of the optic axis as source are given by the following expression
\begin{equation}
 \theta_{m,n}=-\frac{C_m}{D_d}+\frac{1}{D_d}\frac{D_m}{B_m}exp\left(-\frac{(2n+1)\pi}{A_m} \right)\left( 1-\frac{1}{A_m}\frac{D_s}{D_{ds}}\beta \right).
\end{equation}

When the source is exactly behind the lens i.e. when $\beta=0$, the magnification diverges. It corresponds to the tangentical caustic. We get Einstein rings whose radii are given by 
\begin{equation}
  \theta_{Em,n}=\frac{C_m}{D_d}-\frac{1}{D_d}\frac{D_m}{B_m}exp\left(-\frac{(2n+1)\pi}{A_m} \right).
\end{equation}
There are no radial caustics.

Thus we get new infinite set of images and Einstein rings below the critical angle due to the light rays that enter the outer photon sphere just above the peak and get reflected inside the matter region.

The time required for the light to travel from source to observer is given by 
\begin{eqnarray}
 t_m &=& 2\int_{r_0}^{r_2}\frac{V_{m}(r_0)}{f_m(r)\sqrt{V_{m}(r_0)-V_m(r)}}dr  + 2\int_{r_2}^{D_{d}}\frac{V_{m}(r_0)}{f_2(r)\sqrt{V_{m}(r_0)-V_2(r)}}dr \nonumber \\
     &=& 2\int_{r_0}^{r_2}\frac{V_{m}(r_0)}{f_m(r)\sqrt{V_{m}(r_0)-V_m(r)}}dr + 2\int_{0}^{x_{max}} \frac{V_m(r_0)}{f_2(r(x))}T_m(r(x))\frac{1}{\sqrt{V_{m}(r_0)-V_{2}(r(x))}}dx 
     \label{t2t}
\end{eqnarray}
$x_{max}$ which is very close to one corresponds to the $r=D_d$. 
The first term in the first line in the equation above gives finite contribution since the range of integration is finite and further although integrand diverges at $r_0$ the integral is finite since slope of $V_m(r)$ is finite. Whereas second term will diverge, firstly because the integrand is finite at infinity and secondly because $(V_{m}(r_0)-V_2(r))$ goes to zero at outer photon sphere $r=r_{ph2}$ if $r_0$ is close to $r_m$. 

The expression for $t_m$ can be written as 
\begin{equation}
 t_m= 2\int_{r_0}^{r_2}\frac{V_{m}(r_0)}{f_m(r)\sqrt{V_{m}(r_0)-V_m(r)}}dr +
     2\int_{0}^{x_{max}}G_{m}(r_0,x)g_{m}(r_0,x)dx,
     \label{t22m}
\end{equation}
where $G_{m}(r_0,x)$ and $g_{m}(r_0,x)$ are given by 
\begin{eqnarray}
 G_{m}(r_0,x)&=&  \frac{V_m(r_0)}{f_2(r(x))}T_m(r(x)) \nonumber \\
g_{m}(r_0,x) &=&  \frac{1}{\sqrt{V_{m}(r_0)-V_{2}(r(x))}}. 
\end{eqnarray}

We now write down the term that is sourced by infinity. While writing down the expression below we make use of Eq.\ref{rx}, Eq.\ref{tmm}, Eq.\ref{t22m}.
\begin{eqnarray}
 I_{Dm,1}&=&\int_{0}^{1}G_{m}(r_m,x)g_{m}(r_m,x=1)dx \nonumber \\
 &=& \frac{4M_2}{1-f_2(r_2)}\int_{0}^{1} \frac{1}{(1-x)^2( f_2(r_2)+ (1-f_2(r_2))x )}dx.
\end{eqnarray}
Evaluating the integral above we get 
\begin{equation}
 I_{Dm,1}=2D_{d}-4M_2\log{\left( \frac{r_2}{D_d}\right)}-6M_2-4M_2\log(f_2(r_2)),
\end{equation}
where $D_d$ is distance from lens to observer or source.

We now try to capture arising from the photon sphere when $r_0$ is very close to $r_{ph2}$.
\begin{equation}
 I_{Dm,2}=2 \int_{0}^{1}G_{m}(r_m,r_{ph2})\frac{1}{\sqrt{\alpha_m+\beta_m (x-x_{ph2})^2}}dx,
\end{equation}
where $\alpha_m$, $\beta_m$ were defined earlier in Eq.\ref{alm}. We compute the integral above and Taylor exapnd the result about outer photon sphere. We get
\begin{equation}
 I_{Dm,2}=-\bar{A}_m\log{\left( 1-\frac{r_0}{r_m} \right)}+\bar{B}_m+O\left(1-\frac{r_0}{r_m} \right),
\end{equation}
where 
\begin{eqnarray}
 \bar{A}_m &=& \frac{2G_m\left(r_m,x=x_{ph2} \right)}{\sqrt{\beta_m(r_{ph2})}}, \nonumber \\
 \bar{B}_m &=& \frac{2G_m\left(r_m,x=x_{ph2} \right)}{\sqrt{\beta_m(r_{ph2})}}\log\left( \frac{4 x_{ph2}(1-x_{ph2})\beta_m(r_{ph2})}{-V_m^{'}(r_m) r_m} \right).
 \end{eqnarray}
 
 Regular part of the integral is given by 
 \begin{eqnarray}
  I_{Rm,2}(r_0)&=& 2\int_{r_0}^{r_2}\frac{V_{m}(r_0)}{f_m(r)\sqrt{V_{m}(r_0)-V_m(r)}}dr+2\int_{0}^{x_{max}} G_{m}(r_0,x)g_{2}(r_0,x)dx \nonumber \\
               && -2\int_{0}^{x_{max}} G_{m}(r_{ph2},x=1)g_{2}(r_0,x)dx -2\int_{0}^{1} G_{m}(r_m,x_{ph2})\frac{1}{\sqrt{\alpha_{m}+\beta_{m}(x-x_{ph2})^2}}dx \nonumber \\ &=&  I_{Rm,2}(r_m) +O\left(1-\frac{r_0}{r_m} \right).
 \end{eqnarray}
 
 Combining regular and divergent terms we get 
 \begin{eqnarray}
  t_m (r_0)&=& 2D_d-4M_2\log{\left( \frac{r_2}{D_d} \right)}-6M_2 -4M_2 \log\left(f_2(r_2) \right) \nonumber \\ && -\bar{A}_m\log\left(1-\frac{r_0}{r_m} \right)+\bar{B}_m + I_{Rm,2}(r_0)
  \label{mf1}
 \end{eqnarray}
 
 Not for all values of reflection point $r_0$ images are formed, i.e. light from the source getting reflected back would reach the observer.
 For image to form the following condition should be met. 
 \begin{equation}
  \hat{\alpha}_{m,n}=-A_{m}\log{\left(1-\frac{r_0}{r_m}\right)}+\tilde{B}_m+I_{Rm} - \pi \approx 2n\pi.
  \label{mf2}
 \end{equation}
 Combining the equations Eq.\ref{mf1}, Eq.\ref{mf2} we get time required for the formation of $n^{th}$ image.
 \begin{equation}
  t_{m,n}=2D_d-4M_2\log{\frac{r_2}{D_d}}+\frac{\bar{A}_m}{A_m}(2n+1)\pi+ I_{m}^{'},
 \end{equation}
where
\begin{equation}
 I_{m}^{'}=-6M_2-4M_{2}\log{\left(f_{m}(r_m)\right)}-\frac{\bar{A}_m}{A_m} \tilde{B}_m -\frac{\bar{A}_m}{A_m} I_{Rm}+\bar{B}_m+I_{Rm,2}(r_m)
\end{equation}
Time delay between $p^{th}$ and $q^{th}$ order images is given by 
\begin{equation}
 t_{m,p}-t_{m,q}=2 \frac{\bar{A}_m}{A_m}(p-q)\pi.
\end{equation}
This is the time delay between the same set of images i.e. images formed due to the light rays that get reflected in the matter region.

\begin{figure}
\begin{center}
\includegraphics[width=1\textwidth]{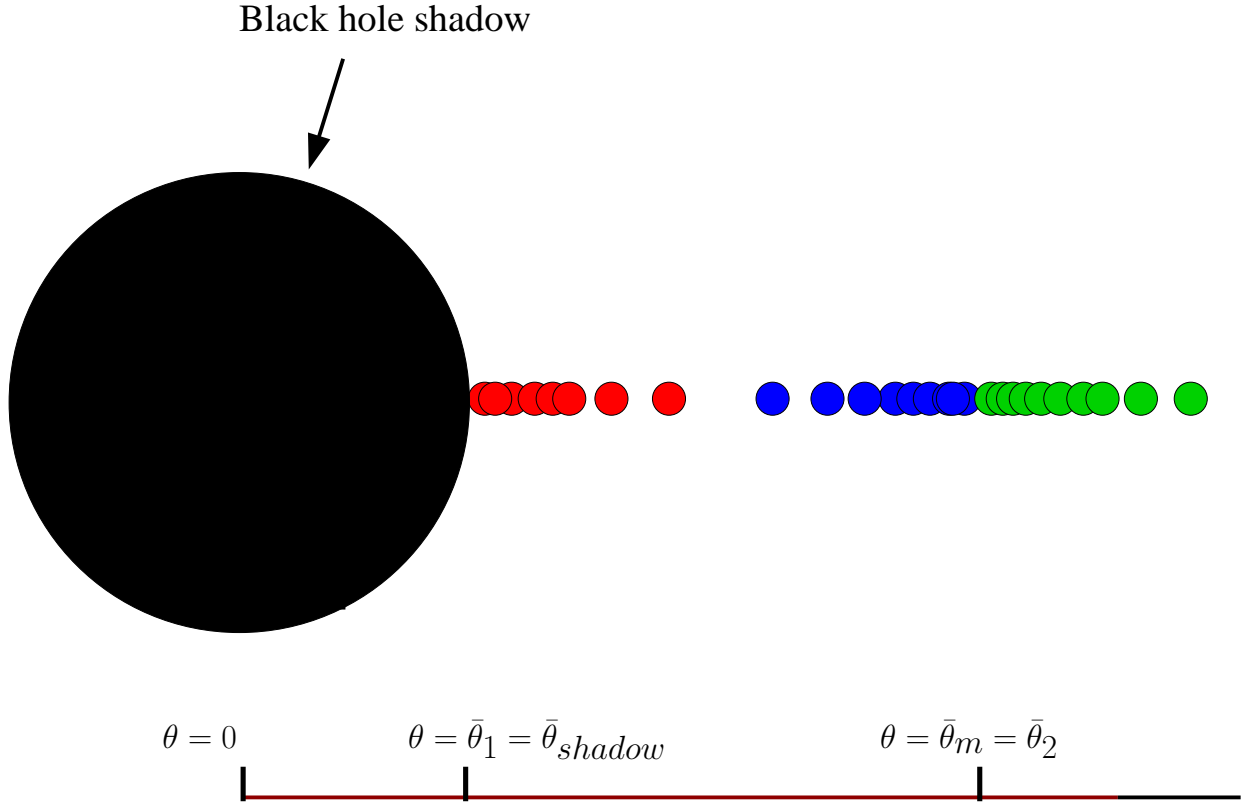}
\caption{\label{img}
In this figure we depict the pattern of images formed by the black hole surrounded by matter. Images drawn as green dots are formed due to the light rays that get reflected just outside the outer photon sphere. Images lie beyond and are cluttered near the critical angle $\theta=\bar{\theta}_2$. Images depicted by blue dots are formed due to the light rays that enter outer photon sphere just above the peak and get reflected back in matter region. They are cluttered just below the critical angle $\theta=\bar{\theta}_2=\bar{\theta}_m$. Images due to the light rays that get reflected just outside the inner photon sphere are depicted by red dots. They are cluttered just outside $\theta=\bar{\theta}_1$, where $\bar{\theta}_1 < \bar{\theta}_2$. No images occur below $\theta=\bar{\theta}_{shadow}=\bar{\theta}_{1}$. This region is shadow of the black hole. 
}
\end{center}
\end{figure}

\section{Images due to the light rays that get reflected just outside the inner photon sphere}

In this section we focus on the light rays that enter outer photon sphere, pass over to the inner region through the matter and get reflected back just outside the inner photon sphere located at $r_{ph1}=3M_1$. The value of $\frac{1}{b^2}$ is slightly below the maximum of the effective potential $V_{1m}=\frac{1}{27M_1^2}$ and therefore both $\dot{r}$ and $\ddot{r}$ tend to zero in this region. Consequently light spends a lot of time in this region while it circles the black hole with finite angular velocity resulting into large deflection of light. 

The deflection angle is given by the expression 
\begin{equation}
 \hat\alpha_{1}=I_1-\pi,
\end{equation}
where $I_1$ is given by 
\begin{equation}
 I_1=2\int_{r_0}^{r_1}\frac{1}{r^2\sqrt{V_1(r_0)-V_1(r)}}dr +  
 2\int_{r_1}^{r_2}\frac{1}{r^2\sqrt{V_1(r_0)-V_m(r)}}dr +
 2\int_{r_2}^{\infty}\frac{1}{r^2\sqrt{V_1(r_0)-V_2(r)}}dr.
\end{equation}
Here $r_0$ corresponds to the radial location where light turns back in the vicinity of inner photon sphere. First, second and third integrals correspond to the contribution to the deflection angle in the inner region , matter region and outer region respectively. The second integral is finite because it is computed over a finite range of radial coordinate and integrand is finite. Third integral is computed over the infinite range of radial coordinates. We an easily verify that it is finite by redefining new coordinate $\omega=\frac{f_2(r)-f_2(r_2)}{1-f_2(r_2)}$ and writing integral in terms of it since the range of integration becomes finite and also the integrand is finite everywhere.

In order to compute the first integral we make the coordinate change 
\begin{equation}
 z=\frac{f_1(r)-f_1(r_0)}{f_1(r_1)-f_1(r_0)}.
\end{equation}
$z$ is a monotonically increasing function of $r$. It varies from $0$ to $1$ as $r$ varies from $r_0$ to $r_1$. We can invert it to write down 
\begin{equation}
 r=\frac{1}{\frac{(1-z)}{r_0}-\frac{z}{r_1}}.
\end{equation}

We define function $T_1(r_0,r)$ as 
\begin{equation}
T_1(r_0,r)=\frac{dr}{dz}=\frac{f_1(r_1)-f_1(r_0)}{f^{'}_{1}(r)},
\end{equation}
which is the derivative of two radial coordinates. The integral $I_1$ can now be written as 
\begin{equation}
 I_1=\int_{0}^{1}\frac{F_1(r_0,r(z))}{\sqrt{V_1(r_0)-V_1(r(z))}}dz +  
 2\int_{r_1}^{r_2}\frac{1}{r^2\sqrt{V_1(r_0)-V_m(r)}}dr +
 2\int_{r_2}^{\infty}\frac{1}{r^2\sqrt{V_1(r_0)-V_2(r)}}dr,
\end{equation}
where $F_1(r_0,r(z))=\frac{2T_1(r_0,r(z))}{r(z)^2}$.

We can carry out analysis similar to the one in section V, and write down the divergent part of the integral as 
\begin{eqnarray}
 I_{D1} &=& F_1\left(r_{ph1},r_{ph1}\right)\int_{0}^{1} \frac{dz}{\sqrt{\alpha_1 z+\beta_1 z^2}}  \nonumber \\  &=& -A_1 \log{\left(1-\frac{r_{ph1}}{r_0}\right)}+\tilde{B}_{1} +O \left( 1-\frac{r_{ph1}}{r_0}\right),
\end{eqnarray}

with $A_1$ and $\tilde{B}_1$ as 
\begin{eqnarray}
 A_1 &=& \frac{F_1(r_{ph1},r_{ph1})}{\sqrt{\beta_1(r_{ph1})}} \nonumber \\
 \tilde{B}_1 &=& \frac{F_1(r_{ph1},r_{ph1})}{\sqrt{\beta_1(r_{ph1})}} 
 \log{\left( \frac{2T_1(r_{ph1},r_{ph1})}{r_{ph1}} \right)},
\end{eqnarray}
where
\begin{eqnarray} 
\alpha_1(r_0)&=&-T_1(r_{ph1},r_{ph1})V_1^{''}(r_{ph1})r_{ph1}\left(1-\frac{r_{ph1}}{r_0}\right) + O\left(\left(1-\frac{r_{ph1}}{r_0}\right)^2 \right)
\nonumber \\
\beta_1(r_0)&=&-\frac{1}{2}T_1^2(r_{ph1},r_{ph1})V_1^{''}(r_{ph1})
+O \left( 1-\frac{r_{ph1}}{r_0}\right).
\end{eqnarray}

Regular part of integral is given by
\begin{eqnarray}
  I_{R1}(r_0) &=&\int_{0}^{1} \left( \frac{F_1(r_0,r(z))}{\sqrt{V_1(r_0)-V_1(r(z))}} -
  \frac{F_1(r_{ph1},r_{ph1})}{\sqrt{\alpha_1 z+ \beta_1 z^2}} \right)dz \nonumber \\ 
  &&+  
 2\int_{r_1}^{r_2}\frac{1}{r^2\sqrt{V_1(r_0)-V_m(r)}}dr +
 2\int_{r_2}^{\infty}\frac{1}{r^2\sqrt{V_1(r_0)-V_2(r)}}dr \nonumber \\
 \nonumber \\
 &=&   I_{R1}(r_{ph1})+ O\left( \left(1-\frac{r_{ph1}}{r_0}\right) \right)
\end{eqnarray}

Combining together regular and divergent part we get an expression for deflection angle 
\begin{equation}
 \hat\alpha_{1}= -A_1 \log{ \left(B_1 \left(1-\frac{r_{ph1}}{r_0}\right) \right)} -\pi + O\left(\left(1-\frac{r_{ph1}}{r_0}\right)\right) ,
\end{equation}
where
\begin{equation}
 B_1=exp\left(-\frac{\tilde{B}_1+I_{R1}}{A_1} \right).
\end{equation}

We Taylor expand impact parameter around the inner photon sphere
\begin{equation}
 b_2=C_1+D_1\left( 1-\frac{r_{ph1}}{r_0} \right)^2+O\left(\left(1-\frac{r_{ph1}}{r_0}\right)^3\right),
\end{equation}
where 
\begin{eqnarray}
 C_1 &=& \frac{1}{\sqrt{V_1(r_{ph_1})}} \nonumber \\
 D_1 &=& -\frac{1}{4}\frac{\left(V_1^{''}(r_{ph1})\right) }{V_1^{\frac{3}{2}}(r_{ph1})}r_{ph1}^2
\end{eqnarray}

Location of images is given by 
\begin{equation}
 \theta_{1,n}=\frac{C_1}{D_d}+\frac{1}{D_d}\frac{D_1}{B_1^2}exp\left(-\frac{2(2n+1)\pi}{A_1} \right)\left( 1+\frac{2}{A_1}\frac{D_s}{D_{ds}}\beta\right).
\end{equation}
We get infinitely many images which lie towards right in the vicinity of critical angle given by
\begin{equation}
 \bar{\theta}_1=\frac{C_1}{D_d}=\frac{3\sqrt{3}M_1}{D_d}.
\end{equation}
Note that this critical angle is smaller than the critical angle encountered earlier, i.e.
\begin{equation}
 \bar{\theta}_1 < \bar{\theta}_2=\bar{\theta}_m.
\end{equation}
Magnification of images is given by 
\begin{equation}
 \mu_{1,n}=\frac{2}{\beta}\frac{D_s}{D_{ds}D_d^2}\frac{D_1}{A_1 B_1^2}\left(\frac{D_1}{B_1^2}exp\left( -\frac{4(2n+1)\pi}{A_1} \right) + C_1 exp\left( -\frac{2(2n+1)\pi}{A_1} \right)\right).
\end{equation}
These are the results for the images that lie on the same side as of the optic axis as source. Location of images that lie on the opposite side is given by 
\begin{equation}
 \bar{\theta}_{1,n}=-\frac{C_1}{D_d}+\frac{1}{D_d} \frac{D_1}{B_1^2}exp\left(-\frac{2(2n+1)\pi}{A_1} \right)\left( 1-\frac{2}{A_1}\frac{D_s}{D_{ds}}\beta\right).
\end{equation}

There is no radial caustic. Tangential caustic occurs when the source is located exactly behind the lens, i.e. when $\beta=0$ and we get Einstein rings whose angular radius is given by 
\begin{equation}
 \theta_{E1,n}=\frac{C_1}{D_d}+\frac{D_1}{B_1^2}\frac{1}{D_d}exp\left(-\frac{2(2n+1)\pi}{A_1} \right).
\end{equation}
All the rings are located slightly above the critical angle $ \bar{\theta}_1$.

Thus we get very interesting pattern for images and Einstein rings. We get three distinct set of infinite images and rings as depicted in Fig.6 and Fig.7. For the isolated black hole we get only one set of images and rings. Thus gravitational lensing signature for matter distribution considered here is very peculiar and distinct. 

The time required for light to travel from source to observer is given by 
\begin{equation}
\begin{split}
t_1 =&2\int_{r_0}^{r_1} \frac{\sqrt{V_1(r_0)}}{f_1(r)\sqrt{V_1(r_0)-V_1(r)}}\dd{r} +2\int_{r_1}^{r_2} \frac{\sqrt{V_1(r_0)}}{f_m(r)\sqrt{V_1(r_0)-V_m(r)}}\dd{r} \\ &+2\int_{r_2}^{D_d} \frac{\sqrt{V_1(r_0)}}{f_2(r)\sqrt{V_1(r_0)-V_2(r)}}\dd{r} 
\\
=& 2\int_{0}^{1} \frac{\sqrt{V_1(r_0)}}{f_1(r(z))}T_1(r_0,r(z))\frac{1}{\sqrt{V_1(r_0)-V_1(r(z))}}\dd{z} +2\int_{r_1}^{r_2} \frac{\sqrt{V_1(r_0)}}{f_m(r)\sqrt{V_1(r_0)-V_m(r)}}\dd{r} \\ &+2\int_{0}^{x_{max}} \frac{\sqrt{V_1(r_0)}}{f_2(r(x))}T_m(r(x))\frac{1}{\sqrt{V_1(r_0)-V_2(r(x))}}\dd{x}. \\
\end{split}
\end{equation}
$x_{max} $ is  very close to one and corresponds to $ r=D_d $. The first term in the diverges because the term $ \sqrt{V_1(r_0)-V_1(r)} $ goes to zero at $r=r_0$ which is in the proximity of the inner photon sphere $ r=r_{ph1} $, while the second term is finite. The third term diverges as the integrand is finite  at infinity.
 The expression for $ t_2 $ can be rewritten as 
\begin{equation}
 t_1 = 2\int_{0}^{1}  G_{1}(r_0,z) g_{1}(r_0,z)\dd{z} +2\int_{r_1}^{r_2} \frac{\sqrt{V_1(r_0)}}{f_m(r)\sqrt{V_1(r_0)-V_m(r)}}\dd{r}  +2\int_{0}^{x_{max}}G_{2}(r_0,x) g_{2}(r_0,x)\dd{x}. 
\end{equation}
 where
 \begin{eqnarray}
 G_{1}(r_0,z)&=&  \frac{\sqrt{V_1(r_0)}}{f_1(r(z))}T_1(r_0,r(z))\nonumber \\
 g_{1}(r_0,z) &=&  \frac{1}{\sqrt{V_1(r_0)-V_1(r(z))}} \\
 G_{2}(r_0,x)&=&  \frac{\sqrt{V_1(r_0)}}{f_2(r)}T_m(r(x)) \nonumber \\
 g_{2}(r_0,x) &=& \frac{1}{\sqrt{V_1(r_0)-V_2(r(x))}}. 
 \end{eqnarray}
 
We now write down the term that diverges when $ r_0 $ is very close to the inner photon sphere $ r_{ph1} $.
\begin{equation}
I_{D1,1} = 2\int_{0}^{1}  G_{1}(r_{ph1},z=0)\frac{1} {\sqrt{\alpha_1 z+ \beta_1 z^2}}\dd{z}  
\end{equation}
By following similar calculations as done in earlier sections, the divergent part of the integral can be calculated  as
\begin{equation}
	I_{D1,1}  = -\bar{A}_1 \log{\left( 1- \frac{r_{ph1}}{r_0}\right) } + \bar{B}_1 + O \left(  1- \frac{r_{ph1}}{r_0}\right), 
\end{equation} 
where
\begin{eqnarray}
\bar{A}_1 &=& \frac{2G_1\left(r_{ph1},z=0 \right)}{\sqrt{\beta_1(r_{ph1})}}, \nonumber \\
\bar{B}_1 &=& \frac{2G_1\left(r_{ph1},z=0 \right)}{\sqrt{\beta_1(r_{ph1})}} \log(\frac{2T_1(r_{ph1},r_{ph1})}{r_{ph1}}).
\end{eqnarray}

Now, we try to find the contribution to divergence due to the third integral in the expression for $ t $.
\begin{equation}
	I_{D1,2} = 2\int_{0}^{x_{max}}G_{2}(r_0,x) g_{1}(r_0,x=x_{max})\dd{x}.
\end{equation}
Evaluating the above integral we get 
\begin{equation}
I_{D1,2}=2D_{d}-4M_2\log{\left( \frac{3M_2}{D_d}\right)}-6M_2-4M_2\log(f_2(r_{ph2})).
\end{equation}
Now, the regular part of the integral is given by,
\begin{equation}
\begin{split}
I_{R1,2}(r_0) &= 2\int_{0}^{1}  G_{1}(r_0,z) g_{1}(r_0,z)\dd{z} +2\int_{r_1}^{r_2} \frac{\sqrt{V_1(r_0)}}{f_m(r)\sqrt{V_1(r_0)-V_m(r)}}\dd{r} \\
& +2\int_{0}^{x_{max}}G_{2}(r_0,x) g_{2}(r_0,x)\dd{x}
-2\int_{0}^{1}G_{1}(r_0,z) \frac{1}{\sqrt{\alpha_1 z+ \beta_1 z^2}}\dd{z} \\
& -2\int_{0}^{x_{max}}  G_{2}(r_{ph1},x=x_{max}) g_{2}(r_0,x)\dd{x}\\
&= I_{R1,2}(r_{ph1}) +O\left( 1- \frac{r_{ph1}}{r_0}\right).
\end{split}
\end{equation}
Combining the regular and divergent part, we get
 \begin{eqnarray}
t_1 (r_0)&=& 2D_d-4M_2\log{\left( \frac{3M_2}{D_d} \right)}-6M_2 -4M_2 \log\left(f_2(r_{ph2}) \right) \nonumber \\ && -\bar{A}_1\log\left(1-\frac{r_{ph1}}{r_0} \right)+\bar{B}_1 + I_{R1,2}(r_0).
\end{eqnarray}
The images are not formed for all values $ r_0 $, but only  for the values of $ r_0 $ close to the inner photon sphere. The condition to be met for the formation of image is 
 \begin{equation}
\hat{\alpha}_{1,n}=-A_{1}\log{\left(1-\frac{r_{ph1}}{r_0}\right)}+\tilde{B}_1+I_{R1} -\pi \approx 2n\pi.
\end{equation}
Combining the above two equations,  we get time required for the formation of $n^{th}$ image.
\begin{equation}
t_{1,n}=2D_d-4M_2\log\left( {\frac{3M_2}{D_d}}\right) +\frac{\bar{A}_1}{A_1}(2n+1)\pi+ I_{1}^{'},
\end{equation}
where
\begin{equation}
I_{1}^{'}=-6M_2-4M_{2}\log{\left(f_{2}(r_{ph2})\right)}-\frac{\bar{A}_1}{A_1} \tilde{B}_1 -\frac{\bar{A}_1}{A_1} I_{R1}+\bar{B}_1+I_{R1,2}(r_{ph1})
\end{equation}
Time delay between $p^{th}$ and $q^{th}$ order images is given by 
\begin{equation}
t_{1,p}-t_{1,q}=2 \frac{\bar{A}_1}{A_1}(p-q)\pi.
\end{equation}
This is the time delay between the same set of images i.e. images formed due to the light rays that get reflected close to the inner photon sphere.

\begin{figure}
\begin{center}
\includegraphics[width=0.9\textwidth]{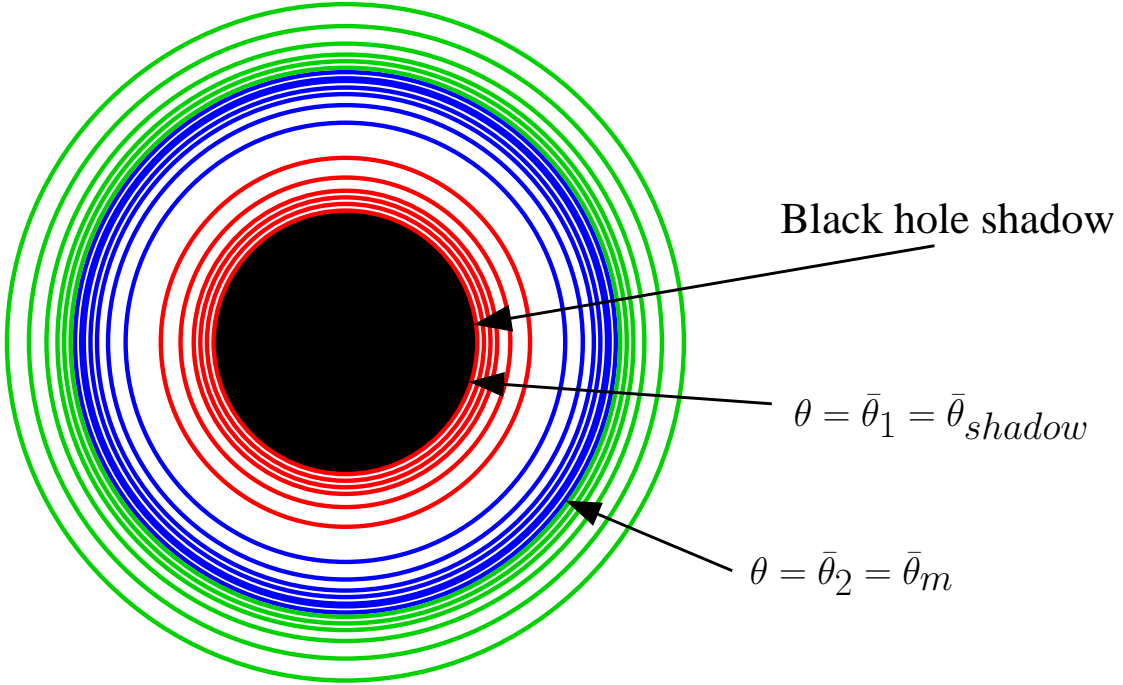}
\caption{\label{eir}
In this figure we depict the pattern of Einstein rings formed by the black hole surrounded by matter when the source is exactly behind the lens. In the middle we get region with radius $\theta=\bar{\theta}_{shadow}$ where no Einstein rings appear. This is called black hole shadow. Red circles depict Einstein rings formed due to the light rays that get reflected back just outside the inner photon sphere. Rings lie just outside and are cluttered around $\theta=\bar{\theta}_1=\bar{\theta}_{shadow}$. Blue rings are formed due to the light rays that enter outer photon sphere just above the peak and get reflected in the matter region. These lie below and are cluttered around $\theta=\bar{\theta}_m$. Green rings are formed due to the light rays that get reflected just outside the outer photon sphere. They lie outside and are cluttered around the critical angle $\theta=\bar{\theta}_1=\bar{\theta}_m$
}
\end{center}
\end{figure}

\section{Light rays which enter inner photon sphere and black hole shadow}
We considered three distinct scenarios depending on the location of turning point for the light rays and showed that we get three distinct set of infinitely many images and Einstein rings. We now consider the fourth category of the light rays. The rays for which $\frac{1}{b^2}$ is larger compared to the potential maximum at inner photon sphere $V_{1m}=\frac{1}{27M_1^2}$. These light rays enter the inner photon sphere, beyond which they  do not encounter any other potential barrier and hence admit no turning point. Thus they are destined to enter the event horizon and are engulfed by the black hole. Thus the ingoing light rays do not reach infinity. Thus we would see a dark region below critical angle $\bar{\theta}_1$ circular in shape. This is known as shadow of the black hole. Shadow has an angular size given by expression 
\begin{equation}
 \bar{\theta}_{shadow}=\bar{\theta}_1=\frac{3\sqrt{3}M_1}{D_d}.
\end{equation}
The shadow is depicted in the Fig.6 , Fig.7. 

Since size of the shadow is dictated by the inner photon sphere it will be proportional to mass $M_1$. So if we observe the shadow of the configuration considered here, its angular size will allow us to infer the mass $M_1$. The situation would be very different if we try to infer the mass associated with configuration based on the observation of motion of the distant object. 
The mass inferred from such a consideration will be $M_2$. This results into the situation where mass inferred from two different methods would yield conflicting answers. This would imply the existence of matter surrounding black hole. Thus before arriving at radical conclusions such as breakdown of general relativity or new physics, we should consider conservative scenarios such as existence of matter, perhaps dark matter, in the vicinity of black hole.

\section{Image location, Magnification and time delay between the most prominent images}

It is quite clear from the expression for the magnification that it decays exponentialy as the the number of turns light takes around the black hole goes on increasing. Generally, in case of a single black hole, the first images would be relevant from the point of view of observation and other infinite set of higher order images can be ignored since they are highly demagnified. In presence of matter as we have shown earlier three set of infinitely many images would occur depending on their turning points. In this section, we enlist the most prominent images that would be of consequence from the point of observations.

For this calculation we work in the units where $ M_1=1 $. We take $ M_2=1.5 $. So the photon spheres occur at $r_{ph1}=3$ and $r_{ph2}=4.5$. The matter distribution extends from $r_1=3.5$ to $r_2=3.7$. We use a mass function in the matter region that is used in \cite{Konoplya}. The distance between source and lens as well as observer and lens is taken to be $D_d=D_{ds}=1000$. The source location is $\beta=10^{-2}$.

\begin{table}[h!]
	\begin{tabular}{|c|c| c| c|} 
		\hline
		& $ \theta_2 $ & $ \mu_2 $ & $ t_2 $ \\ [0.5ex] 
		\hline
		n=1 & 0.00780411 &  $1.51 \times 10^{-5} $ & 2099.58\\ 
		\hline
		n=2 & 0.00779424 &  $2.82 \times 10^{-8} $ & 2148.55\\ 
		\hline
	\end{tabular}
	\caption{In this table we list the the image location, magnification and the time taken for light rays to reach the observer from the source to observer for light rays that get reflected near the outer photon sphere. Only the first image is prominent as the magnification of the other images are much smaller.}
	\label{outerps}
\end{table}

\begin{table}[h!]
	\begin{tabular}{|c|c| c| c|} 
		\hline
		& $ \theta_m $ & $ \mu_m $ & $ t_m $ \\ [0.5ex] 
		\hline
		n=1 & 0.00764038 &  $-1.16 \times 10^{-4} $ & 2312.89\\ 
		\hline
		n=2 & 0.00778755 &  $- 5.12 \times 10^{-5} $ & 2313.58\\ 
		\hline
		n=3 & 0.00779391 &  $- 2.22 \times 10^{-7} $ & 2314.28\\ 
		\hline
	\end{tabular}
	\caption{In this table we list the the image location, magnification and the time taken for light ray to reach the observer from the source in case of light rays that get reflected in the matter region. The first two image has magnification comparable to the magnification of images due the first set of light rays (reflected near outer photon sphere). The magnification of images beyond this are minuscule. }
	\label{matterregion}
\end{table}

\begin{table}[h!]
	\begin{tabular}{|c|c| c| } 
		\hline
		& $ \theta_1 $ & $ \mu_1 $ \\ [0.5ex] 
		\hline
		n=1 & 0.00519615777 &  $4.94 \times 10^{-9} $ \\ 
		\hline
		n=2 & 0.00519615016 &  $1.01 \times 10^{-10} $\\ 
		\hline
	\end{tabular}
	\caption{In this table we list the the image location and magnification for light ray to reach the observer from the source in case of light rays that get reflected close to the inner photonsphere. Images have magnification that is four orders of magnitude smaller than that of lighter image. }
	\label{innerps}
\end{table}

In the absence of any matter, if we have a black hole with mass $M_2=1.5$ as seen from outside, the brightest image will be the first order image. It will occur at $\theta_{2,1}=0.00780411$ and will have magnification $\mu_{2,1}=1.51\cross 10^{-5}$. In the presence of the matter the brightest image will be the first order image due to the light rays turn back inside matter region and occur at $\theta_{m,1}=0.00764038$ and it will have magnification
$\mu_{m,1}=1.16\cross 10^{-4}$ which is ten times brighter. Second order image will occur at $\theta_{m,2}=0.00778755$ and will have magnification $5.12 \cross 10^{-5}$ which will be five times brighter. All other images will have smaller magnification (See Tables I,II,III.) which we ignore. Thus in the presence of matter we will have two images brighter than what we would have in case of a single isolated black holes.

If the source is variable, then the variability will be reflected onto the three images at different time. The time taken by the rays getting reflected back outside outer photon sphere and constituting first order is $t_{2,1}=2099.58$, onto which the variability will be reflected initially. Then it would then be reflected onto the first order image for the the light rays that get reflected in the matter region after time $t_{m,1}-t_{2,1}=213.31$.It will be almost 10 times brighter. Following which it will be reflected onto the second order image with light turning back in middle region after $t_{m,2}-t_{m,1}=0.69$. It will be dimmer by the factor of 2.

The apperance of the two new prominient images due to the light rays reflecting back inside the matter region, as opposed to just one prominient image in case of a an isolated single black hole and peculiar time delay between the images are the smoking gun signature of the presence of matter surrounding the black hole.

\section{Summary and Conclusion}

With observational probes such as Event Horizon telescope reaching unprecedented accuracy, we are living in the era where direct observation of black holes is possible. It will allow us to study the astrophysical environment and various physical processes that occur in the vicinity of black holes at the center of our and neighboring galaxies. It could be the case that significant amount of matter, which could be the conglomeration of dark matter for instance, could be present in the vicinity of black holes.
In the work we try to consider such a situation and made an attempt to understand whether matter distribution around the central black hole could significantly affect the observations, specifically focusing on the gravitational lensing, pattern of images, Einstein rings and shadow cast by such a configuration. For this purpose instead of getting into detailed modeling of such a scenario we deal with a very simple toy-model for the matter distribution and show that it can affect the gravitational lensing in a crucial way and leave its imprint on the pattern of images, Einstein rings as well on the shadow of the black hole. As it turns out photon sphere plays a very important role in the gravitational lensing investigations.

We consider a Scharzschild black hole with mass $M_1$ located at center. Its photon sphere is located at $r_{ph1}=3M_1$. The mass outside the matter distribution is $M_2$, the photon sphere associated with which is located at $r_{ph2}=3M_2$. We assume that the matter lies entirely between radii $r=r_1$ and $r=r_2$  which in turn lie between the two photon spheres.
We assume that the mass function $m(r)$ which is constant for Schwarzschild metric, varies between $r=r_1$ and $r=r_2$, i.e. in the region where matter is located. It is a monotonically increasing function and interpolates between the mass in the inner region $M_1$ and $M_2$ which is the mass in the outer region as we go from $r_1$ to $r_2$. We make a further assumption that mass function $m(r)$ is chosen in such a way that effective potential decreases monotonically and does not admit any peak, i.e. no unstable photon sphere in the region where matter is located. As a consequence of which effective potential for the radial motion of light rays admits an interesting pattern.    
It starts from a zero value at event horizon located at $r=2M_1$, goes on increasing, admits a maximum at inner photon sphere $r=r_{ph1}=3M_1$. It then goes on decreasing all the way upto outer boundary of the photon sphere $r=r_2$. It then increases and admits a maximum at the location of outer photon sphere $r=r_{ph2}=3M_2$ and then goes to zero at we reach infinity. Since the maximum of the potential is given by the expression $V=\frac{1}{27M^2}$, the height of maximum at outer photon sphere is smaller compared to the height at the inner photon sphere. Thus we encountered different scenarios. Depending on the value of impact parameter $b$ we can have initially ingoing light rays that turn back outside the outer photon sphere, light rays which enter outer photon sphere above the peak at maximum and encounter potential barrier in the matter region and get reflected back and the light rays which enter the inner region passing over the outer and matter region and get reflected back outside the inner photon sphere. If the reflection point is very close to the photon sphere or if light ray passes just above the peak of the potential, both radial velocity $\dot{r}$ and rate of change of radial velocity $\ddot{r}$ approach zero. Thus light ray spends a lot of time close to the photon sphere. Since it has finite angular speed it will go around the black hole large number of times resulting into extremely large deflection angle, which shows divergence in the limit when the proximity to the photon sphere tends to zero. Thus we get light rays which go around the black hole once, twice, and all the way upto infinitely many times resulting into infinitely many relativistic images. If the source is exactly behind the black hole we get Einstein rings. Thus in this case we would get three distinct sets of infinitely many images and Einstein rings corresponding to three different scenarios described earlier depending on
where light turns back. This is a very peculiar and distinct feature. This is in contrast with the situation where we have only one set of infinite images and Einstein rings in case of the single isolated black hole. Since the relativistic images are highly demagnified and brightness decreases exponentially as the number of turns light takes around black hole goes on increasing, we try to identify most prominent images. We infer that three images are prominently visible as compared all other infinitely many higher order images as compared to just one image for single isolated black hole. 

The relativistic images are de-magnified exponentially as the number of turns light takes around the black hole increases. Hence very few out of the infinitely many images are relevant from the point of view of observations. We compute magnification and time delays in a specific case and infer that three images are relevant from the point of view of observation. This includes two images from the light rays which enter outer photon sphere and get reflected back in the matter region and one image from the light rays that get reflected back outside outer photon sphere. This is quite different from the case of single isolated black hole where only the first relativistic image will be relevant from the point of view of observations. This is very peculiar feature would be a smoking gun signature of matter distribution around black hole. 

If light enters the inner photon sphere, it does not encounter a potential barrier. Thus it inevitably enters the event horizon and is engulfed by the black hole. Thus we get a dark circular patch in the middle devoid of any image or Einstein ring. It is known as shadow of the black hole. In this case size of the shadow of the black hole, i.e. angular radius of the dark patch is dictated by the inner photon sphere and is thus proportional to mass $M_1$. If we observe the shadow of the black hole, the mass inferred from it will be $M_1$. We can also estimate mass of the whole configuration by monitoring the motion of the distant objects in the gravitational field. Mass inferred from such a method would yield $M_2$. So there is a discripency in the inference about the mass made from two different observations. It will provide yet another evidence in favor of mass distribution around the black hole.

By invoking a simplistic toy model of matter distribution we demonstrated that pattern of images and Einstein rings as well as shadow measurement displays a very interesting trend quite distinct from that of a single isolated black hole. In future we intend to study more realistic scenarios with better modeling of the matter distribution around black holes.

\end{document}